\documentclass{emulateapj}
\newcommand{\hideindraft}[1]{}
\usepackage{amsmath}
\usepackage{graphicx}  
\usepackage{hyperref}
\usepackage{color}

\newcommand\unit[1]{\, {\rm #1}}

\newcommand\eqnbreak[1]{ \nonumber \\ &#1& }

\newcommand{\bi}{\begin{itemize}}
\newcommand{\ei}{\end{itemize}}
\newcommand{\be}{\begin{enumerate}}
\newcommand{\ee}{\end{enumerate}}

\newcommand\editremark[1]{ {\color{red} #1}}

\newcommand{\er}{\ensuremath{\epsilon_{\rm radio}}}
\newcommand{\eedd}{\ensuremath{\epsilon_{\rm Edd}}}
\newcommand{\fgeo}{\ensuremath{f_{\rm geo}}}

\begin{document}

\title{Blindly Detecting Orbital modulations of jets from merging Supermassive Black Holes }
\author{ R.~O'Shaughnessy}
\email{oshaughn@gravity.phys.uwm.edu}
\author{D.~L.~Kaplan}
\affil{Physics Dept., U. of Wisconsin - Milwaukee, Milwaukee
  WI 53211}
\email{kaplan@uwm.edu}
\author{A. Sesana}
\affil{Albert Einstein Institute, Am Muhlenberg 1 D-14476 Golm,
  Germany; and Center for Gravitational Wave Physics, The Pennsylvania
  State University, University Park, PA 16802}
\email{alberto.sesana@aei.mpg.de}
\and
\author{A. Kamble}
\affil{Physics Dept., U. of Wisconsin - Milwaukee, Milwaukee
  WI 53211}

\slugcomment{Draft \today}

\begin{abstract}

In the last few years before merger, supermassive black hole binaries will rapidly inspiral and precess in a magnetic field
imposed by   a surrounding circumbinary disk.   Multiple simulations suggest this relative motion will convert some
of the local energy to a Poynting-dominated outflow,  with a luminosity $\sim 10^{43}\unit{erg\,s}^{-1}\,(B/10^4 G)^2(M/10^8
M_\odot)^2 (v/0.4 c)^2$, some of which may emerge as synchrotron
emission at frequencies near 1\,GHz where current and planned
wide-field radio surveys will operate. 
On top of a secular increase in power (and $v$) on the gravitational wave inspiral timescale, orbital motion will produce significant, detectable modulations, both  on orbital periods and (if black hole spins are
not aligned with the binary's total angular momenta) spin-orbit precession timescales.  
Because the gravitational wave merger time increases rapidly with separation, 
we find vast numbers of
these transients are ubiquitously predicted, unless explicitly ruled out (by low efficiency $\epsilon$) or
obscured (by accretion geometry $f_{geo}$).
If the fraction of Poynting flux converted to radio emission times the fraction of lines of sight accessible
$f_{geo}$ is sufficiently large
($f_{geo} \epsilon > 2\times  10^{-4}$ for a 1 year orbital period), at least one event is  accessible to future blind surveys at a
nominal $10^4\unit{deg}^2$ with $0.5\unit{mJy}$ sensitivity. 
Our procedure generalizes to other flux-limited surveys designed to investigate  EM signatures associated with  many
modulations produced by merging SMBH binaries.
\end{abstract}


\section{Introduction}
Merging supermassive black hole binaries should naturally possess a circumbinary accretion disk
whose MRI-driven turbulence generates and imposes  a substantial external magnetic field.  
Recent simulations suggest that supermassive black holes (SMBHs)  moving through an imposed magnetic field should produce
a Poynting-dominated outflow, consisting of a jet \citet{pll10} and   diffuse 
emission \citet{luciano-newjet}.    Thus, even in the absence of accretion, during their last years before merger, SMBH binaries should generally have faint emission, modulated by their orbital motion.
Though the outflow (henceforth ``jet'') power increases during the inspiral, ending in a bright flare, as discussed in
\cite{bbhpaper}, 
such flares are too faint or too rare to easily detect, unless almost all the Poynting flux is efficiently converted
to low-frequency radiation.
On the other hand, because SMBH binaries spiral in very slowly through gravitational wave emission, each merger flare is
preceded by a long phase during which the jet is modulated and growing, at only slightly reduced efficiency.
These modulations should be easily accessible to future radio surveys.

In this paper we calculate how frequently modulations will occur and how often they can be detected.  
Our paper adopts and extends the assumptions used in \cite{bbhpaper} for the  jet power versus binary
masses and for fiducial SMBH binary merger rates.
In \S \ref{sec:rates} we sum over all SMBH binaries, to determine the number of binaries on our past light cone with
detectable modulation.
In \S \ref{sec:jet} we describe  how long, how frequent, and how bright modulations from each SMBH binary should be.
Finally,  \S \ref{sec:detect:Advanced} we discuss, in order to detect the 
 signature presented here, targeted EM and GW surveys must overcome limitations of intrinsic AGN variability or small limiting distances, respectively.
We also briefly explain how the many distinctive variations in the predicted
light curve will distinguish this source from other candidate modulation.


To date, most $1\unit{GHz}$ radio transient surveys have surveyed at most a few thousand square degrees with rms
sensitivities of $\sim\unit{ mJy}$ \citep{ofb+11}, with many having smaller area, less sensitivity, and sporadic
sampling.   Upcoming surveys will be larger, more regular, and more sensitive. 
  In particular, the VAST (Variables and Slow Transients;
\citet{mc+11}) project using the Australian Square Kilometer Array Pathfinder (ASKAP; \citet{jbb+07}) will survey
roughly $10^4\,\unit{deg}^2$ daily down to a nominal sensitivity of $0.5 \unit{mJy}$.   We adopt these parameters to
motivate our discussion, keeping in mind that survey plans often change and  that unknown factors  such as efficiencies
or obscurations strongly impact our results.   Rather than fix these choices, we leave in scalings, so interested parties can make their own predictions. 

\subsection{Context }
\label{sec:context}
Several mechanisms for precursor, prompt, and delayed  electromagnetic signatures of SMBH merger have been proposed, 
 including Poynting-dominated jets \citep{bbhpaper,2010Sci...329..927P,luciano-newjet}; disk emission powered by
post-merger perturbations and shocks \citep{2008ApJ...684..835S,2009ApJ...700..859O,2009PhRvD..80b4012M,2010MNRAS.401.2021R}; the viscous-driven refilling of the post-merger
disk \citep{2005ApJ...622L..93M,2010ApJ...714..404T}; accretion of a fossil inner disk driven by the smaller body's inspiral \citep{2010MNRAS.407.2007C}; increased tidal disruption
rates \citep{2009ApJ...697L.149C,clus-smbh-TidalDisruptionRates-WeggBode2011,clus-smbh-TidalDisruptionRates-StoneLoeb2011} and perturbations
of galactic-center stellar cores; and  direct modulation of a close circumbinary disk
\citep{2009CQGra..26i4032H,smbh-bin-merger-EMC-Disk-BodeBogdanovic-2011,2011MNRAS-Roedig}; see \citet{review-Schnittman2010} and
references therein.
%
These merging binaries will also produce gravitational waves potentially  accessible to pulsar timing arrays and LISA
 \citep{whitepaper-PulsarTiming1,whitepaper-PulsarTiming2,2009MNRAS.394.2255S,2002MNRAS.331..805H,2007MNRAS.377.1711S}.
However measured, the SMBH merger rate would provide invaluable direct constraints on the merger rates of galaxies
 \citep[versus direct methods, as discussed in][and references therein]{2009MNRAS.396.2345B} and the  
formation and evolution of supermassive black holes \citep[see,\,e.g.][and references therein]{2010AARv..18..279V}.  
%
Depending on the identification epoch,  mechanism, and followup observations, direct detection of merging SMBH binaries
could provide infomation about accretion, stellar dynamics, the evolution of galaxies
\citep{2010AARv..18..279V,whitepaper-CoordinatedScience}.

A number of binary SMBH   
are known or suspected \citep[e.g.,][]{komossa06,rtz+06,ssb+10,bs10}.  
While many dual AGN candidates (e.g., 3C75 and NGC 6240)  have been discovered with $\unit{kpc}$-scale-separation, very few candidate SMBH binaries are
known with $<10\unit{pc}$ separation.
Objects with shifted broad line regions like  SDSS J092712.65+294344.0 \citep{2003ApJ...582L..15K} or
J153636.22+044127.0  \citep{2009Natur.458...53B} may be SMBH binaries; see, e.g.,
\cite{2009ApJ...697..288B,2009MNRAS.398L..73D} and references therein.
   \citet{2006ApJ...646...49R} demonstrate O402+379 has two resolved  cores with a
projected $7\unit{pc}$ separation.  More speculatively, OJ287 has precessing jets \citep{2011ApJ...729...26M} and outbursts with an approximately 12
year period \citep{1988ApJ...325..628S}.  These modulations have been fit both by binary motion \citep{2008Natur.452..851V} and by a warped accretion disk \citep[see,\,e.g.][]{katz97}.
Since most  SMBH binaries 
are too far away to  be resolved 
by VLBI, future candidates will likely be found by offset emission lines or modulated spectral features
\citep[e.g.,][]{2009ApJ...698..956C,2010ApJ...725..249S,2011arXiv1106.1180T}. 
For example, recently, \citet{smbh-Eracleous-SurveyForClose-2011} found 14 quasars with offset H$\beta$ emission lines, with offsets that shifted significantly between
widely separated observations.  These deviations could arise from line-of-sight orbital acceleration in an SMBH binary. 
%


The search for binary SMBH is complicated by AGN variability.  This variability is largely not periodic; detecting such periodic behavior in a
radio light curve would be a strong indication of an inspiralling SMBH
pair \citep[e.g.,][]{komossa06}.   

Our study complements the previous model-neutral investigation by   \cite{smbh-bin-HKM2009}, who also both calculate the number of SMBH
binaries per unit orbital period and the conditions under which (optical) surveys could detect their variability; see,
e.g., their Figure 9.    They examined both gravitational-wave-dominated and disk-driven inspiral. 
By contrast, in this paper we study the gravitational-wave-dominated phase only; adopt a concrete emission model; employ
a state-of-the art population of merging SMBH binaries distributed over cosmic time; and  address how AGN background
variability and jet beaming would impact survey performance.  We also discuss how gravitational wave searches with pulsar
timing naturally complement EM surveys for massive, comparable-mass SMBH binaries.

\section{Population of modulated outflows from merging binaries}
\label{sec:rates}









Following \cite{2010Sci...329..927P} and \cite{bbhpaper}, we assume each merging binary black hole has a
Poynting-dominated outflow, even in the absence of accretion.\footnote{\label{foot:Main}On physical grounds we expect $L_{jet}\propto B^2
  M^2 q^2 v^2$, based on the black hole area, the energy density of the magnetic field, and the relative orbital
  velocity of the hole and ambient field.   Self-consistent MRI amplification already produces a large enough    $B$ field in the disk 
  \citep[see,\,e.g.][]{2006PhRvL..97v1103P} to power the instantaneous outflow assumed above.    Following
  \cite{2009ApJ...707..428B},  \cite{2008ApJ...677.1221R}, and references therein, we anticipate small loops will escape into the
  circumbinary region, expand rapidly in the absence of confining material, and form a fluctuating but partially  coherent large-scale
  field in the interior that threads the circumbinary region and the hole.  Ordered ambient flux can be advected inward,
   increasing the ordered  field near the horizon.  Neither process  requires matter
  accretion.  Still higher magnetic fluxes can be achieved in the presence of accreting matter, a small
  amount of which will flow inward from the nonrelaxed inner edge of the circumbinary disk.
A detailed discussion of $B$ field evolution and flux advection onto the horizon is beyond the scope of this paper.  
Rather than adopt a model for $B$ fields near the hole, as  described
  in \cite{bbhpaper} we conservatively adopt a smaller magnetic field, such that the maximum jet luminosity just prior
  to merger is the Eddington luminosity if $\eedd=1$. 
}   For convenience, we scale the outflow's instantaneous luminosity to a  fraction $\eedd$ of the  Eddington luminosity at the merger event:
\begin{eqnarray}
\label{eq:def:JetPower}  
L_{flare} = \eedd q^2 (v/v_{max})^2 L_{edd} 
\end{eqnarray}
where $q=m_{2}/m_{1}<1$ is the binary's mass  ratio and $v=\beta c$ is the binary's coordinate relative circular
velocity ($|\partial_t (\vec{r}_2-\vec{r}_1)|$.
\cite{2010Sci...329..927P} find that most of the emission is
trapped into collimated jets, with $\eedd \approx  0.002$.   Using a refined calculation of the outgoing flux,
\citet{luciano-newjet}  calculate that the jets, while present, are associated with  a much brighter diffuse quadrupolar
emission; in total,  \citet{luciano-newjet} estimate the same fields correspond to   $\eedd \approx 0.02$.
%
%
In either scenario,  the Poynting-dominated outflow will naturally interact with the strong ambient  magnetic field and
coronal plasma to produce radiation. 
  Lacking a detailed emission model, we assume a fraction $\er$ of the beam power is
converted to radio as $F_\nu\propto \er L/\nu$ and absorb all details of spectral model,  K-correction, and most issues
pertaining to anisotropic emission\footnote{Later we will explicitly account for how orientation-dependent obscuration ($f_{geo}$) or
  beaming ($f_{beam}$) impacts the number versus flux distribution.} into $\er$.   A survey with a flux threshold $F_{\nu,min}$ will therefore be sensitive to all binaries  inside
a luminosity distance $d_{\rm Edd}(v,M,q)$:  
\begin{eqnarray}
\label{eq:Scalefac}
d_{\rm Edd} &\simeq& 
     \sqrt{L/4\pi F_{\rm min}} 
\eqnbreak{\simeq} 14.2\unit{Gpc} (\beta/\beta_{max}) \sqrt{\frac{q^2(M/10^6
  M_\odot)(\epsilon/0.002)}{(F_{\nu,{\rm min}}/\unit{mJy})(\nu/{\rm
      GHz})}} \\
\epsilon&\equiv& \epsilon_{radio}\eedd
\end{eqnarray}
where $M$ is the total mass of the binary.  All key results and figures are presented as functions of $\epsilon$, to
allow efficient scaling to any physically motivated values of $\eedd,\er$.

The jet power gradually increases over time as the binary spirals inward  towards merger due to the influence of
gravitational radiation,  with circular velocity increasing as
\begin{eqnarray}
\label{eq:Inspiral}
\frac{d\beta}{dt} &=& \frac{c^3}{G}\frac{96}{15}\frac{\eta}{M}\beta^9
\end{eqnarray}
where the symmetric mass ratio $\eta=m_1 m_2/M^2 = q/(1+q)^2$.
As each outflow is tied to a single black hole, we anticipate modulation at the orbital period 
\begin{eqnarray}
\label{eq:def:Period}
P &=&  \frac{G}{c^3} 2\pi M \beta^{-3}
\end{eqnarray}
Finally, for black holes with spin $S=\chi m^2$ misaligned with the orbital angular momentum, the orbital plane should
precess, leading to modulations on a timescale of order \citep{acst94} 
\begin{eqnarray}
\frac{2\pi}{T_{\rm prec}} &\simeq &\frac{G}{c^2} \frac{J}{r^3}(2+3 q) 
  \eqnbreak{\simeq }
\label{eq:def:Prec}
\frac{G}{c^2}\times \begin{cases}
 \frac{S_1}{r^3}(2+3 q)  \simeq 
  2 \frac{|\chi| }{Mc } \beta^6   & \text{(late)} \\
 \frac{L}{r^3}(2+3 q)  \simeq
   2 \frac{q}{M c} \beta^5   & \text{(early)}
\end{cases}
\end{eqnarray}
In these expressions we differentiate between early in the inspiral, when the orbital angular momentum is much larger
than the black hole spin, and late in the inspiral, when $L\lesssim S_1$.   Though the latter case occurs only in the
last few orbits before merger with significant spin-orbit coupling, it ensures a  significant change in the direction of $\hat{L}$ during each precession
period and provides the best opportunity for jet modulation.   On the contrary, early in the evolution, the orbital
plane precesses through a small angle $\simeq {\cal O}(S/L)$; while many cycles occur, only for exceptionally bright jets with
well-understood variability could we ascribe small  modulations  to precession.
Spin-orbit misalignment  may not occur  in gas-rich mergers, as accretion may align spins to the orbital plane \citep{2007ApJ...661L.147B}.

\begin{figure}
\includegraphics[width=\columnwidth]{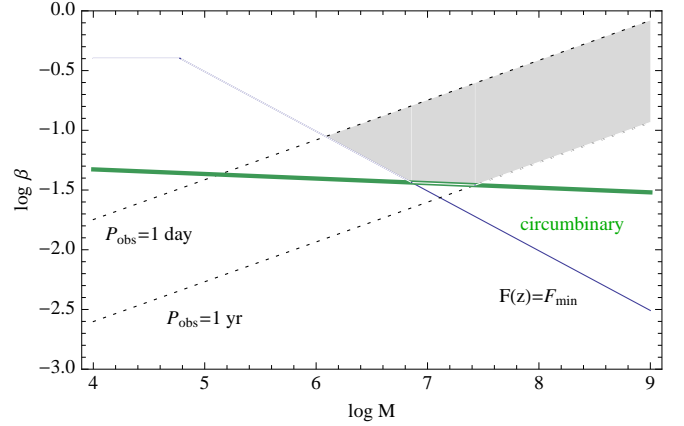}
\caption{\label{fig:TimescaleDiagram} \textbf{Critical orbital velocities versus mass}: For an equal-mass binary of mass
  $M$ at $z=0.6$, a plot 
  of orbital velocity versus mass.  The two dotted upward-sloping lines are the velocities associated with 1 year (bottom) and 1
  day (top) orbital periods;
 a one year survey with one day cadence will be roughly sensitive to this range of time periods.
  The downward-sloping blue line shows the minimum velocity  $\beta_c,$ above which the jet is bright enough to be seen at $z=0.6$,
  assuming $\epsilon=0.002$ and $F_{min}=1\unit{mJy}$; this curve depends strongly on source distance, efficiency, and
  survey flux limit.
  Finally, the thick green line is   $\beta_*$, the velocity at which the binary separates from the circumbinary disk
  [Eq. \ref{eq:BetaCrit:SimpleScaling}].  The shaded region is bounded by $\beta_\pm$ and corresponds to the range of
  velocities where a binary is both isolated and detectable as a periodic source.  The time $\Delta T$ a binary spends
  evolving through this region is $\Delta T = T(\beta_+)-T(\beta_-)$ for $T$ given by Eq. \ref{eq:def:Tmerge}.
}
\end{figure}

Given these timescales and a birthrate $\partial_t n$ per unit comoving volume and time, we can determine the number of
SMBH binaries on our past light cone whose modulation timescales lie in a desired range and whose jets are bright enough
to detect:
\begin{eqnarray}
N_I &=& \int dV_c \left[\int_{\beta_-(z)}^{\beta_+(z)} d\beta \frac{\partial_t n}{d\beta/dt} \right] \nonumber \\
\label{eq:SnapshotNumber}
&=& \int dV_c \partial_t n  \left[T(\beta_+)-T(\beta_-) \right]
\end{eqnarray}
where $dV_c$ is the comoving volume element, where $T(v)$ is the time until merger for a binary of velocity $c\beta$
\begin{eqnarray}
\label{eq:def:Tmerge}
T(\beta) =  - \frac{G}{c^3} \frac{5}{256} \frac{M}{\eta} \beta^{-8}
\end{eqnarray}
and  where the redshift-dependent expressions $v_{\pm}$ are defined using the smallest interval satisfying (a) the desired time-velocity relationship; (b) the constraint that
$d<d_{Edd}$;  (c) that $T(\beta_{\pm})$ are less than 
the 
time since decoupling from the circumbinary disk $\simeq {\cal O}(1)\unit{Myr}\, q^{7/13} (M/10^8 M_\odot)^{17/13}$ 
\citep[see,\,e.g.,][and references therein]{2008ApJ...684..835S,2009ApJ...700..859O}.\footnote{For brevity, in our
  expressions we assume an accretion rate $\dot{m}=0.3 \dot{M}_{edd}$  and disk viscosity parameter $\alpha=0.3$,
  following \citet{2007MNRAS.376.1740K}.
}
For example, the total number of binaries whose observed periods lie in $[P_{obs,-},P_{obs,+}]$ follows from 
\begin{subequations}
\label{eq:BetaCrit:SimpleScaling}
\begin{eqnarray}
\beta_c &\equiv& \beta_{max} \sqrt{4\pi d_L^2F_{min}\Delta \nu /L_{edd} \epsilon q^2} \\
\beta_* &\equiv& T^{-1}(
0.86 \unit{Myr}\times q^{7/13} (M/10^8 M_\odot)^{17/13}) \\
\beta_{\pm} &=& \text{min}[ \beta_{max}, \text{max}[P^{-1}(P_{obs,{\mp}}/(1+z)), \beta_c,\beta_*]] 
\end{eqnarray}
\end{subequations}
where $T^{-1},P^{-1}$ are the inverse functions of Eqs. \ref{eq:def:Period},\ref{eq:def:Tmerge}; a similar expression with $P\rightarrow T_{\rm prec}$ holds for
precession.   
 This expression distinguishes between the source-frame period $P$ and the observed period
  $P_{obs}=P(1+z)$.
Figure \ref{fig:TimescaleDiagram} illustrates how these four critical surfaces define a range of orbital velocity $\beta$ consistent
with maximum source distance, maximum lifetime, and allowed period range.
For example, circumbinary disks limit the orbital period of \emph{decoupled, GW-inspiral-dominated} binaries to $P \lesssim
3\unit{yr} q^{15/26} (M/10^8 M_\odot)^{29/26}$.  Similarly, circumbinary disks a priori limit the number of decoupled BH binaries  to
less than $\simeq 1\unit{Myr}$ times the all-sky merger rate:  less than $10^7$, or $<240\unit{deg}^{-2}$.
%
This limit is reached only for very sensitive surveys that probe all black holes throughout the universe.  As
implied by   Figure \ref{fig:TimescaleDiagram}, for typical surveys and for the vast
majority of decoupled binaries at moderate redshift,  the
  flux limit or period limit bounds the space of detectable binaries; the circumbinary disk plays no role.


Following  \cite{bbhpaper}, we calculate the distribution of modulated flux at the earth using a distribution of merger
rates as a function of black hole mass and cosmic time.  The assembly of SMBHs is reconstructed through dedicated Monte Carlo
merger simulations which are framed in the hierarchical structure
formation paradigm.  Briefly, these models evolve the BH population
starting from BH ``seeds,'' through accretion episodes triggered by
galaxy mergers, and include the dynamical evolution of SMBH-SMBH
binaries.  The SMBH population is consistent with observational
constraints, e.g.,  the luminosity function of quasars at $1<z<6$, the
$M-\sigma$ relation and the BH mass density at $z=0$
\citep{2003ApJ...582..559V,2008MNRAS.383.1079V,2010MNRAS.409.1022V}.
To illustrate the situation, we adopt two of the  fiducial
merger distributions as used in \citet{abb+09}.  The two models used
here are representative of two proposed seeding scenarios (see
\citealt{2011PhRvD..83d4036S} for further discussion).  In the
notation of \citet{abb+09}, we compare models LE and SE, where S
versus L refers to the seed size -- large or small -- and E refers to
``efficient'' accretion. The models described in that paper are
representative of a range of plausible SMBH growth scenarios.  
As with uncertainties in $\er$, we attempt to make our conclusions
robust to specific merger assumptions.

Any reasonably-large sky patch should contain many bright SMBH binary jets whose plausible modulation timescales are
within reach: minutes to months.   Precessing sources, though the most unambiguous sources of modulation, are rare: only for the most
optimistic efficiencies ($\epsilon \simeq 1$) will a typical-scale survey ($10^4 \unit{deg}^2$) have sufficiently bright
 precessing jet in its sky patch [Figure   \ref{fig:EddingtonB:AllSkyBrightJets:Precession}].  On the contrary, many
 binaries have orbital periods between minutes and months  and are close enough to produce bright jets
[Figure   \ref{fig:EddingtonB:AllSkyBrightJets:OrbitPeriod}].
The slow nature of gravitational wave losses insures most binaries are discovered near the widest orbits.  By contrast, 
though the largest and closest binaries are most likely to be detected, any binary not inconsistent with the flux limit
of the survey and that can plausibly be produced at that mass and redshift has a reasonable chance to be  recovered [Figure  \ref{fig:EddingtonB:AllSkyBrightJets:OrbitPeriod:MassRedshift}].
On the one hand,  each value of 
$F/\epsilon$ defines a  minimum mass versus redshift $M_c(z)  \propto F/\epsilon$ below which no emission is
visible even at merger;\footnote{This mass-redshift boundary is defined by  $\epsilon L_{edd}/4\pi d_L^2 = F_{min}$ and would be
  attained only at merger of a comparable-mass binary. }
 these limiting curves are shown in Figure  \ref{fig:EddingtonB:AllSkyBrightJets:OrbitPeriod:MassRedshift} for several
 choices of $F/\epsilon$.
On the other hand, at each redshift the merger process does not efficiently produce SMBHs above a critical mass.  In our
two merger trees, this maximum mass $M_{mgr}(z)$ is approximately 
\begin{eqnarray}
\label{eq:AlbertoMax}
M_{mgr}(z) \simeq 10^{10-z/2.5} M_\odot 
\end{eqnarray}

As a concrete example, in Figure  \ref{fig:EddingtonB:AllSkyBrightJets:OrbitPeriod:MassRedshift} we show the mass-redshift
distribution expected for  a fiducial survey with $F=0.5\unit{mJy}$ sensitivity \citep[e.g.][]{mc+11}, adopting   the most optimistic efficiency $\epsilon=1$.
Though concentrated at moderate mass and redshift, the distribution extends throughout the region bounded by the
smallest flux (black) and largest mass (black dotted).\footnote{For numerical reasons, these boundaries also determine the range of
  fluxes we can reliably model.  Our catalog consists of redshift bins  $\Delta z \simeq 0.1$ out to $z\simeq 10$.  For
  fluxes $F/\epsilon < 0.1 \unit{mJy}$, binaries from $z>10$ may contribute significantly.  By contrast, for $F/\epsilon
  \gtrsim 10^4 \unit{Jy}$, the limiting sensitivity passes inside the smallest redshift bins for a potentially
  significant proportion of mergers.
}
Results at other flux sensitivities are qualitatively similar, derived from this figure by suppressing binaries near and beyond the desired flux limit. 
Finally, the  mass-redshift distribution implies a cumulative flux distribution $N(>F/\epsilon)$, with each value of
$N(>F/\epsilon)$ corresponding to the integral in mass and redshift above a specific threshold curve for $F/\epsilon$  (blue).
In particular, this relationship and these contours suggest that for every flux limit $F/\epsilon<10^3\unit{Jy}$,  a
significant fraction of all deteted binaries are always high mass and moderate- to high-redshift sources ($z>1$).  

Relatively few AGN have unobscured lines of sight to their central engines, particularly at radio frequencies.
Adopting a parameter  $f_{\rm geo}$ to characterize the fraction of unobscured lines of sight,\footnote{This parameter
  absorbs all larger-scale obscuration effects, such as any obscuring torus or clouds that appear in AGN unification
  schemes; see  \cite{1995PASP..107..803U}, \cite{2010ApJ...714..561L} and references therein.  Standard AGN unification
schemes require the torus subtend $65^o$, corresponding to a factor  $f_{\rm geo}^{-1}\simeq 2-4$ in the radio \citep{2010ApJ...714..561L}.}   the cumulative distribution of sources
versus flux is $f_{\rm geo} N(>F/\epsilon)$.  Fitting to the brightest sources in Figure \ref{fig:EddingtonB:AllSkyBrightJets:OrbitPeriod}, we estimate
the number of detectable binaries with periods less than $1\unit{yr}$  is between\footnote{At the highest flux, the local universe
dominates.  Based on  Eq. \ref{eq:SnapshotNumber} and
a uniform local merger rate, we
expect that $N(>F/\epsilon)$ at large flux must scale as 
$
N \propto   
M^{5/6}P^{5/3}/(F/\epsilon)^{3/2}
$.} 
\begin{eqnarray}
\label{eq:NumberDefault}
N_{\rm obs}& \simeq& 
10-30  f_{\rm geo}     
 \left( \frac{\epsilon  \unit{Jy}}{F} \right)
\end{eqnarray}
for a moderate range of  flux ($F/\epsilon \in[10^{-2}, 10^3] \unit{Jy}$).
Taking the lower limit, a fiducial survey with $10^4 \unit{deg}^2$ coverage and $0.5\unit{mJy}$ sensitivity
 should have
at least one binary   with bright, detectable jets and observed period less than $1\unit{yr}$ if the
source obscuration and composite radiative efficiency satisfy
\begin{eqnarray}
\label{eq:CriticalEfficiency:OrbitDoppler}
f_{\rm geo} \epsilon \gtrsim  2\times 10^{-4}(F_{min}/0.5\unit{mJy})
\end{eqnarray}
In terms of Figure \ref{fig:EddingtonB:AllSkyBrightJets:OrbitPeriod}, this constraint corresponds to the number of
orders of magnitude between (a) the flux at which the sky coverage limit (dotted) and the blue cumulative intersect and and (b) the survey
flux limit (just off scale to left).
The number of sources in wide orbits is large;  only extremely inefficient conversion of energy to radio can
prevent them from being seen.

Conversely, Figure \ref{fig:EddingtonB:AllSkyBrightJets:Precession} shows only a few binaries are both massive 
enough to produce a detectable radio jet by this mechanism and tight enough to undergo precession on an accessible
timescale.  Empirically Figure \ref{fig:EddingtonB:AllSkyBrightJets:Precession} is approximated by $N(>F)\simeq
10^{-1.8}(F/\epsilon \unit{Jy})^{-2/3}$.  Therefore, even an equally wide  ($10^4 \unit{deg}^2$) but much deeper ($0.05 \unit{mJy}$) survey will have at
least one binary with bright, detectable precessing jets only if jet power is efficiently converted to radio
energy:
\[
f_{\rm geo}^{3/2} \epsilon > 0.2(F_{min}/0.05\unit{mJy}) \quad \text{(precession)}
\]    

For brevity and clarity in this discussion we assume the outflow emits roughly isotropically in all directions.  If instead the same amount of power is beamed exlusively along a fraction
$f_{beam}<1$ of all lines of sight, the number of detectable sources versus flux $N_{beam}$ can be calculated from the
previous expression via correcting for the fraction of the sources pointing towards us and the increased flux emitted
along the remaining lines of sight: 
\begin{eqnarray}
\label{eq:BeamingCorrection}
N_{beam}(>F/\epsilon)  = f_{beam} N\left( > \frac{Ff_{beam}}{\epsilon} \right)
\end{eqnarray}
Ignoring detectors' sensitivity, the fraction of jets pointing towards us and in the circumbinary phase
 is less than the total number of binaries that have decoupled from their circumbinary disk on our past light cone ($\lesssim
 10^7$) times $f_{beam}\equiv \Delta \Omega/4\pi$, the relative solid angle covered by a jet. 
By contrast, for experimentally accessible and observationally interesting flux ranges, the number versus flux trend $N(>F)\propto 1/F$ [Eq. \ref{eq:NumberDefault} and Figure
  \ref{fig:EddingtonB:AllSkyBrightJets:OrbitPeriod}, leading to $N_{beam}$ independent of beaming.\footnote{For surveys
    limited to very low-redshift sources, $N(>F)\propto F^{-3/2}$.  To the extent that all beamed sources are also
    nearby, beaming would indeed increase the detectable number: $N_{beam}\propto f_{beam}^{-1/2}$.}
While strictly true, beam modulation allows much longer orbits to be accessible and thereby can still
significantly increase the number of accessible sources, as described below.
%

\section{Jet modulation }
\label{sec:jet}

Throughout the inspiral, each black hole drives a Poynting-dominated outflow, powered by the strong ambient field
provided by  the circumbinary disk.  As described in \cite{bbhpaper} (cf. their Footnote 1), our
 assumptions correspond to assuming the disk provides a magnetic field such that the jet luminosity near merger
 $L_j(\beta=\beta_{max})\propto B^2 M^2$ is the efficiency times the eddington luminosity ($\eedd L_{edd}$); the corresponding constant $B$ field is
 $B\simeq B_{edd}\simeq 6\times 10^4 \eedd^{1/2}( M/10^8 M_\odot)^{-1/2}G $.
 A detailed treatment
of the outflow kinematics and emission driven by this ideal-MHD jet is beyond the scope of this paper; see, e.g.,
Section 5 of \cite{2011PhRvD..83f4001L} for plausible emission microphysics and   \cite{2010Sci...329..927P} for MHD. However, the large circumbinary field and
coronal plasma provide a natural mechanism for dissipating the outflow, through strong synchroton losses, inverse
compton scattering off of accelerated particles, an ambient medium to shock, and naturally relativistic particle velocity scales.\footnote{For example, the Poynting flux in
  this ideal-MHD jet ($E.B=0$)
  has a strong  electric field $|\vec{E}|\simeq \sqrt{L_j/r^2}\simeq B_{edd}(M/r)$.  Even in the absence of direct
  acceleration, this  field produces strong transient drift currents
  $v\simeq c$.}
   Following  \citet{bbhpaper}, we assume that a fraction $\epsilon_{\rm radio}$ of the jet power is eventually
emitted at radio frequencies, associated with a jet tied to the spin axes of each black hole. 

As the black holes orbit and precess, the flux along our line of sight will be modulated by orbital effects (e.g., bolometric
Doppler boosting, as in 
\cite{2003ApJ...588L.117L}); by precession of the spin axes changing the 
orientation of the jet; 
and other relativistic effects associated with an ultracompact orbit. 
The observable modulation depends sensitively on   jet dynamics and emission mechanisms, all beyond the scope of this
paper.
As an example with  efficient conversion ($\epsilon \lesssim 1$) and simple modulation, in the rest of this paper we optimistically assume Poynting
flux is efficiently re-radiated by a sufficiently dense corona of weakly relativistic electrons as synchrotron radiation, in a short optically thin low
Lorentz factor ``jet''.  In this scenario, modulations at the orbit period arise from Doppler boosting during the orbit of each unresolved
jet, with relative amplitude $\simeq  \beta/\beta_{max}$ times factors of order unity depending on the line of sight, spectrum, and
mass ratio.  
 Modulations at the precession period arises as the jet flow direction
relative to the line of sight rotates, with line-of-sight flux fluctuations of order unity for a low-Lorentz factor jet.
On the contrary, models with highly relativistic outflows have much stronger beaming,  much less efficient conversion of
jet to radio power, and a weak if any tie between the radio modulation timescale and any binary period: few sources will
point towards us, while each will be radio-fainter and more randomly variable.


\hideindraft{
\subsection{Modulation populations}
* all sky number that change by 10\%

* all sky rate of orbital or precession events versus assumed flux 

* cadence-limited

}

\begin{figure}
\includegraphics{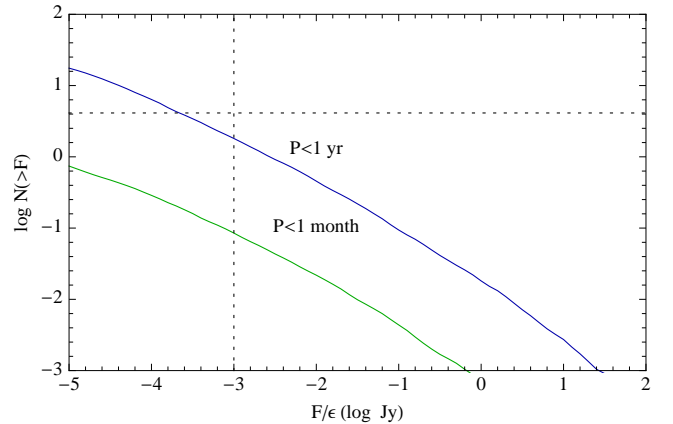}
\caption{\label{fig:EddingtonB:AllSkyBrightJets:Precession}\textbf{Sources with precession modulation: Total number}: Assuming all binaries have significant spin-orbit misalignment,
  the number of sources with precession periods 
$P_{min}=1\unit{min}$ and $P_{max}=1\unit{yr}$ (blue) or  $1\unit{month}$ (green) over the entire sky, with flux above
  $F/\epsilon$. 
 To guide the eye to scales comparable to future radio surveys like VAST \citep{mc+11}, a horizontal dashed line
 corresponds to $1/10^4\unit{deg}^2$; a vertical dashed line corresponds to $1\unit{mJy}$.
\vskip5pt
}
\end{figure}

\begin{figure}
\includegraphics{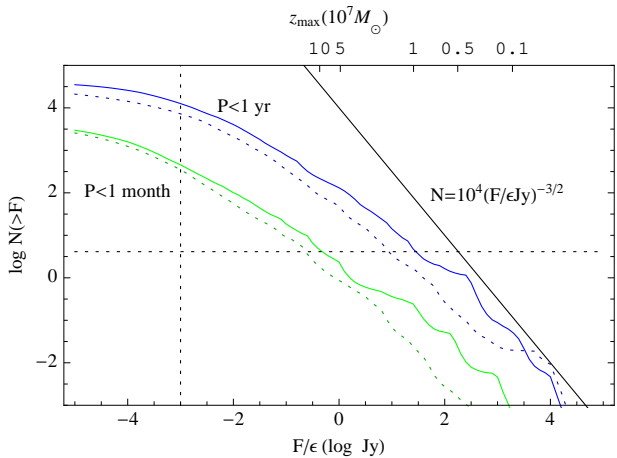}
\caption{\label{fig:EddingtonB:AllSkyBrightJets:OrbitPeriod} \textbf{Sources with orbit period modulation I: Total number}:  At any instant of
  time, the number of decoupled SMBH binaries versus average jet flux, limiting to orbit periods  within
  $P_{min}=1\unit{min}$ and  $P_{max}=1\unit{yr}$ (blue and dotted blue) or  $1\unit{month}$ (green and dotted green) over the entire sky, with flux above
  $F/\epsilon$ for $\epsilon=\er \eedd$ the composite efficiency of the jet, using the SE (small seed, efficient
  accretion; dashed lines) and LE (large seed, efficient accretion; solid lines) model of \citet{2011PhRvD..83d4036S}.
  For example, for an efficiency $\epsilon=10^{-3}$ and target flux sensitivity $1\unit{mJy}$, one to several
  hundred binaries are bright enough to see and have periods less than one year; of order one has a period less than
  1 month.
   In the limit of infinite sensitivity, all rates must be less
  than the all-sky rate ($\simeq 30/\unit{yr}$) times the longest time allowed ($1\unit{Myr}$, corresponding to the time
  since separation from a circumbinary disk). \hideindraft{\editremark{why do all agree at low flux?}}
In the limit of poor sensitivity, the number versus flux scales as $F^{-3/2}$, as usual for the nearby universe.  
For comparison, the solid black line is $10^4 (F/\epsilon Jy)^{-3/2}$.  The top axis shows ticks at $z_{max}$, the
redshift at which $L_{edd}/4\pi d_L^2 = F/\epsilon$ for $M=10^7 M_\odot$; as we assume $\epsilon<1$, no source can be
detected at higher redshift.
\vskip5pt
}
\end{figure}

\begin{figure}
\includegraphics{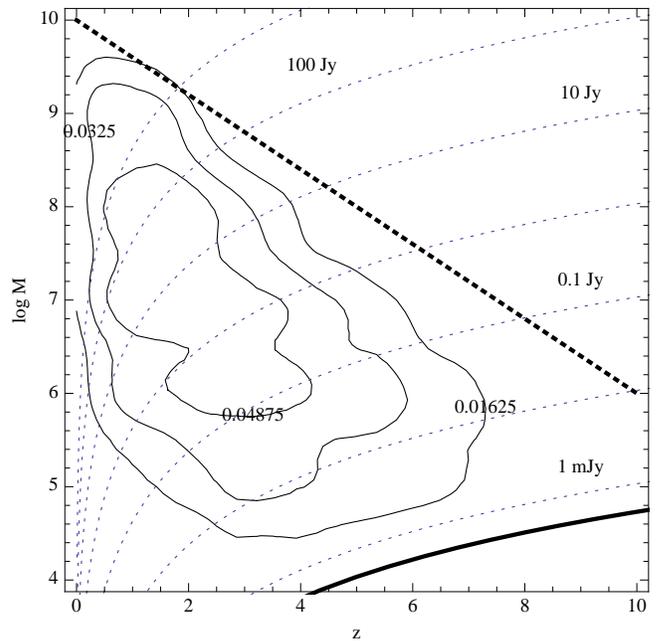}
\caption{\label{fig:EddingtonB:AllSkyBrightJets:OrbitPeriod:MassRedshift} \textbf{Sources with orbit period modulation
    Ib: Mass-redshift distribution}:  Contours of the mass-redshift distribution (solid black) of all binary black hole
  outflows with  $F/\epsilon>=0.5\unit{mJy}$ (thick black curve) and observed periods $P_{\rm obs}$ between $1\unit{yr}$
  and $1\unit{min}$.   This scaled flux limit corresponds to the fiducial
  VAST survey sensitivity ($F=0.5\unit{mJy}$) for the most optimistic conversion of outflow to radio power ($\epsilon=1$).
Contours are shown at $1/4,1/2$, and $3/4$ of the maximum value ($dN/d\log M dz \simeq 0.06$).  The distribution shown
is smoothed, built by convolving a gaussian kernel with the underlying merger tree, with smoothing lengths  $\Delta \log M
\simeq 0.3$ in mass and  ($\Delta z=0.3$) in redshift.
  Also shown are contours of the largest possible (Eddington-limited) flux at a given redshift:
  $F/\epsilon=L_{edd}/4\pi d_L^2 \nu$ for $10^{-3},10^{-2}\ldots 10^3\unit{Jy}$ (dotted blue, bottom to top).  Finally,
  the thick dotted black line is an
  empirical relation for the maximum SMBH binary mass versus redshift (Eq. \ref{eq:AlbertoMax}).
}
\end{figure}

\begin{figure}
\includegraphics{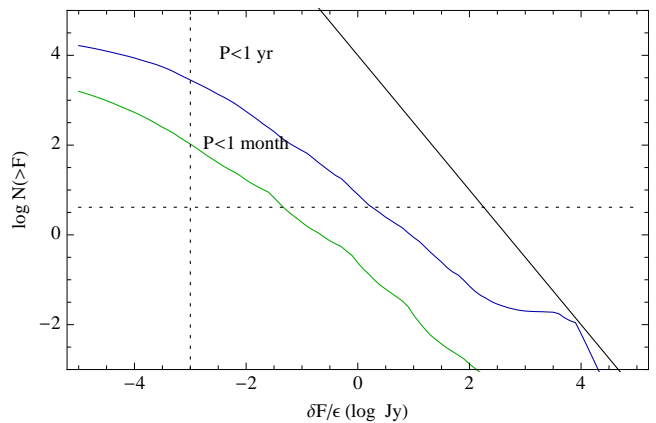}
\caption{\label{fig:EddingtonB:AllSkyBrightJets:OrbitPeriod:Modulated} \textbf{Sources with orbit period modulation II:
    Significant modulation}:  The number of point sources with flux modulation greater than the limiting sensitivity,
  versus that sensitivity.  As Figure \ref{fig:EddingtonB:AllSkyBrightJets:OrbitPeriod:Modulated}, except the number
  is shown versus  $\delta F/\epsilon$, assuming a relative flux modulation $\delta$ per orbit period given by Eq. \ref{eq:RelativeModulation:DopplerBoost}.
}
\end{figure}

To identify modulation, the absolute change in flux from the modulation ($F\times \delta $ for $\delta$ the relative
intensity change)  should be larger than the detector noise.
Equivalently, the per-measurement figure of merit derived from Fourier transforming the recieved flux
\begin{eqnarray}
\label{eq:def:PeriodicDeltaSNR}
\text{SNR} = \delta  \frac{F}{F_{min}}
\end{eqnarray}
should be greater than 1. 
Though realistic surveys will have a higher detection threshold, this factor  can also be absorbed into $\epsilon$.
 For  Doppler boosting at the orbit period, the relative change in power is small for bright
and long-lived jets ($q\simeq 1$), owing
to minimal contrast between the two holes' jets: to order of magnitude the relative power increase 
$\delta$ is [cf. Eq. \ref{eq:def:JetPower}]  
\begin{eqnarray}
\label{eq:RelativeModulation:DopplerBoost}
\delta \simeq \frac{\beta}{\beta_{max}} \frac{1-q^2}{1+q^2}
\end{eqnarray}
  In this case, the analysis of \S \ref{sec:rates} follows, replacing
Eq. (\ref{eq:BetaCrit:SimpleScaling}) with 
\begin{eqnarray}
\beta_{c,rel} &=& \beta_{max} \left(
   \frac{4\pi d_L^2 F_{min}\Delta \nu}{L_{edd} \epsilon q^2} \times \frac{1+q^2}{1-q^2} \right)^{1/3}
\end{eqnarray}
Out of the population of binaries whose jet power is bright enough to be seen and an orbit period in a
testable range [Figure
  \ref{fig:EddingtonB:AllSkyBrightJets:OrbitPeriod}],
Figure \ref{fig:EddingtonB:AllSkyBrightJets:OrbitPeriod:Modulated} shows a not insignificant fraction have 
 jet power bright enough that Doppler boosting could be accessible.
 By contrast, for precession-induced changes in the jet
direction, we expect a sinusoidal relative flux change of order unity: any precessing jet that's bright enough to be seen will have
detectable modulation.   

\begin{figure}
\includegraphics{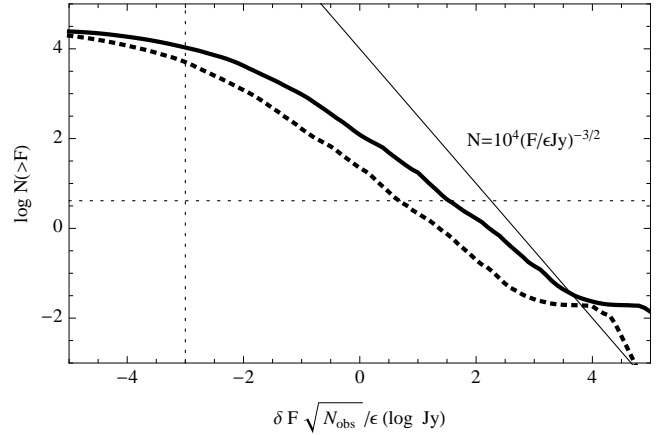}
\caption{\label{fig:Modulation}\textbf{Sources with orbit period modulation III: Stacking significant modulation}:
For the small seed (SE) model,  the cumulative number of sources
  $N(> \delta F \sqrt{N_{obs}}/\epsilon)$ at any instant which could be recovered by a survey with $1/\unit{d}$ (thick) or $7/\unit{yr}$ (dotted),
  assuming one year observing time and stacking all $N_{obs}$ flux measurements into a composite statistic. 
Each curve is directly comparable to the dark blue curve in   Figure
\ref{fig:EddingtonB:AllSkyBrightJets:OrbitPeriod:Modulated}, after shifting the flux scale of the latter by $\sqrt{N_{obs}}$.
}
\end{figure}

To this point we've considered the sensitivity limits for  single measurements.  By folding together multiple cycles we  improve our sensitivity to
faint sources.  A priori, the flux sensitivity increases as the square root of the number of measurements $N_{obs}$; the
figure of merit becomes
\begin{eqnarray}
\label{eq:ImprovementPeriodic:AGNOnly}
{\rm SNR}& =& \delta  \frac{F}{ F_{min}} \times \sqrt{N_{obs}}
\end{eqnarray}
Additionally, the survey duration $T_{obs}$ and cadence $\Delta T = T_{obs}/N_{obs}$ naturally define a minimum and maximum range of periods to which the
survey is sensitive: periods $P_{obs}$ between approximately $2 T_{obs}/N_{obs}$ and $T_{obs}$.  
\hideindraft{Sources with  $P\lesssim \Delta T$ look random and $P\gg T_{obs}$ look constant.}
This subtlety has minimal impact,  as binaries cluster near the longest orbital periods.  As implied by the ratio of
Eqs. \ref{eq:ImprovementPeriodic:AGNOnly} and \ref{eq:def:PeriodicDeltaSNR}, simply rescaling the 
distribution of all binaries whose
peak-to-trough modulations are above the detection threshold [Figure
  \ref{fig:EddingtonB:AllSkyBrightJets:OrbitPeriod:Modulated}] produces the distribution of sources with detectable
periodic flux modulations [Figure \ref{fig:Modulation}].


In the above we have assumed the flux is purely   sinusoidally modulated, suitable to weakly Doppler-boosted synchrotron
emission detected in the radio.  By contrast, a highly relativistic outflow will generally be strongly Doppler boosted,
 tightly beamed, and emit most power preferentially at high energies.   As noted earlier, beamed emission from the rare
 jets pointing towards us could be  detected farther away [Eq. \ref{eq:BeamingCorrection}].
Few binaries  have bright jets and precession periods less than a year  [Figure
  \ref{fig:EddingtonB:AllSkyBrightJets:Precession}].  However,  any transverse structure $\propto \delta \theta$ in the
  narrow jet beam modulates the emitted flux on much shorter timescales  $T_{mod} \simeq \delta \theta/\Omega_p \simeq \sqrt{f_{beam}}T_{prec}$.
The modulation in high-energy
 flux with time  ($\delta(t)$) will generally contain a broad spectrum of frequencies, with frequency content depending on the details of
 the jet beam shape and its orientation relative to the line of sight.
Because many orders of magnitude more binaries have much longer precession periods, these modulations in principle allow
surveys of short cadence to identify sources with long natural periods.  To order of magnitude and focusing only on
bright jets in the local universe, the total number of sources
increases as  (solid angle of jet)$\times$ (inspiral time starting at a  $1\unit{yr}/\delta \theta$ precession period)
$\times  (d(\beta)/\delta \theta)^3$, the increased volume to which a beamed jet can be seen.    Fixing $\beta$ in the center
and final factors by requiring $T_{prec}=1\unit{yr}/\delta \theta \propto \beta^{6}$ [Eq. (\ref{eq:def:Prec})]  and
substituting for $d(\beta)$, \emph{in principle} the number of detectable precessing sources increases
relative to the small-statistics tail Figure \ref{fig:EddingtonB:AllSkyBrightJets:Precession} by a substantial factor
%
\[
\frac{N_{prec,beam}}{N_{prec}} \simeq  \delta \theta^{-3/2}\simeq f_{beam}^{-3/4}
\]
A much less favorable scaling applies at larger distances: following Eq. \ref{eq:NumberDefault}, we expect  $N \propto
\delta\theta^{-1/2}\propto f_{beam}^{-1/4}$.  
In practice,  however, the long-timescale nonperiodic modulations that should arise from slowly precessing, beamed AGN
jets should be difficult to distinguish from background AGN variability, as described in \S~\ref{sec:detect:Advanced} below.


\subsection{Ultrarelativistic outflows}
To this point we have implicitly assumed the radio flux to be weakly modulated on the orbit or precession period,
as expected from prompt emission along a weakly relativistic orbit.   By contrast, if these outflows are similar to
conventional ultrarelativistic outflow models for AGN jets, then time of flight, generation, and reprocessing delays may modify any
intrinsic structure, delaying and distorting the expected pattern from precession or the orbit. 

Slowly precessing ultrarelativistic jets share many qualitative features with more familiar ultrarelativistic outflows
like short GRBs.   In both cases, emission can  be shock-driven, powered by interaction with the ambient medium.
At best, emission from this shock will be delayed from prompt emission by time-of-flight and deceleration delays.
Synchrotron radiation at low frequencies can be self-absorbed; this reprocessing delays and distorts any modulations
imposed by the orbit.  Finally, only observers over a range of angles $\theta_j + 1/\Gamma$ can see the jet (for
$\theta_j$ the jet opening angle, $\Gamma$ the bulk Lorentz factor); it may become visible only at late times. 
Just from outflow dynamics,  the radio power expected from a relativistic outflow should plausibly peak on  timescales of weeks to
months after prompt emission.  

Therefore, ultrarelativistic outflows if present can further bias us towards the longest timescales: sources that
produce the shortest timescales will average away, leaving the multiple-year modulations expected among wide SMBH binaries.\footnote{By contrast, an
  extended epoch of  delayed emission will significantly \emph{enhance} our prospects of detecting single brief events, such
  flares accompanying the merger of SMBH mergers; see Kamble et al 2011.} 

\section{AGN backgrounds and targeted searches   }
\label{sec:detect:Advanced}

To this point we have assumed a blind survey, limited only by background and detector noise.  In practice additional
sources of noise intrinsic and extrinsic to the binary system can severely complicate its detection or identification as
a truly periodic source.  On the contrary, highly relativistic SMBH binaries have unique kinematic and gravitational
wave signatures that help distinguish true binary modulation from a background. 
 Below we discuss two particular sources
of noise  (AGN variabilty and scintillation), and two methods to  verify a particular modulation has binary origin
(multiple timescales and pulsar timing).

\subsection{AGN backgrounds and targeted searches }
The host systems of binaries might be identified by residual accretion-powered emission.
For example, accretion-fueled  AGN jets persist ${\cal O}(\unit{Myr})$ after the black holes stop accreting and separate from
the disk.  Magnetically-powered jets may also already be identified.
Conceivably, a targeted search could much more efficiently identify variability candidates.

On the other hand, 
  AGN jets are well-known to be strongly variable, with flux roughly performing a random walk on long
timescales; see, e.g., \cite{book-Krolik-AGN} and references therein.   Likewise, these magnetically-driven flows 
could be  variable.
%
A periodic modulation
superimposed on a variable jet would be significantly more difficult to detect.

To illuminate the targeting problem, we  parameterize the worst possible scenario:  every merging binary has a bright residual radio jet, emitting at a
significant fraction of the binary's eddington luminosity  ($\epsilon_{\rm
  old} L_{edd}$, with $\epsilon_{old}\gtrsim \epsilon$) and just as variable as active AGN.    For time changes less than a year, AGN fluxes are roughly random
walks in amplitude \citep{1992ApJ...396..469H}.\footnote{We assume no
  correlation between AGN variability and the binary mass.}
Specifically, if $w$ is a white random variable, a prototypical AGN flux and flux power spectrum is
\begin{eqnarray}
F_{AGN}(t) &=& \frac{\epsilon_{\rm old} L_{edd}}{4\pi d_L^2} \int^t w dt \\
S_F(f) &\equiv& \int dt e^{2\pi i f t}\left<F_{AGN}(0)F_{AGN}(t)\right> \nonumber \\
 &=& \delta_{crit}^2 \left(\frac{\epsilon_{\rm old} L_{edd}}{4\pi d_L^2}\right)^2 (f\unit{yr})^{-2} \unit{yr} 
\end{eqnarray}
normalizing to $\delta_{crit}$ relative variations on 1 year timescales.
In this  worst case hypothesis,  for any detectable jet, AGN variability is the dominant noise.  As  both the signal and dominant noise propagate from the source to
our detector,  the figure of merit for detectable sources depends only on ratios of properties at the source:\footnote{Unlike the previous
  case,  the AGN provides a unique noise realization for all detectors; repeated measurements at the same or multiple detectors does not improve our sensitivity.}
\begin{eqnarray}
{\rm SNR} &=& \frac{\delta F}{\sqrt{S_F(1/P)/T}}  \\
  &=&\frac{\epsilon q^2 (\beta/\beta_{max})^3(1-q^2)/(1+q^2)}{\epsilon_{old} \delta_{crit}}
  \sqrt{\frac{T \times 1\unit{yr}}{P^2}} \nonumber \\
 &=&  \frac{\epsilon}{\epsilon_{old} \delta_{crit}} \frac{\sqrt{T  1\unit{yr}}}{P^2} 
\eqnbreak{\times} (2\pi M G/(c\beta_{max})^3)  \frac{q^2(1-q^2)}{1+q^2}
\end{eqnarray}
where for clarity we have replaced $\beta^3$ by the period [Eq. \ref{eq:def:Period}].   

If indeed the residual jet radio power is brighter than  modulation in the magnetically-driven radio emission ($\epsilon
< \epsilon_{old} \delta_{crit}$), the factors in this a priori expression
are potentially unfavorable to targeted searches.  As  black hole
light crossing times ($MG/c^3$)  are less than a few minutes, targeted searches for periodic AGN variability from a jet
superimposed on an existing source  would only be sensitive to short timescales (i.e.,  $P <
\sqrt{1\unit{yr} 500\unit{s}} \simeq 1 \unit{d}$).   Even then, detection could only occur if  the  binary jet and residual jet fluctuations are finely tuned
to nearly the same flux (i.e., $\epsilon \simeq \epsilon_{old} \delta$).

On the other hand,  this
Poynting-dominated outflow isn't driven by accretion and is therefore likely far more stable than accretion-driven
jets.  Additionally, any residual jets should have  dimmed and expanded in the thousands to million years since
accretion terminated.  Assuming the transverse crossing time sets the  characteristic AGN jet variability timescale, a
not-face-on residual jet of age $t_j$ should have variability timescales $\gtrsim \theta_j t_j$: as the jet ages, it
loses the capacity to undergo rapid changes.  As detailed treatment of this kind of residual AGN jet background is
beyond the scope of this paper, we can neither endorse nor rule out the possibility that periodic variations might be
accessible on top of likely jet variability backgrounds.

\subsection{Scintillation}
Emission from distant binaries  propagates through plasma in their host, our galaxy, and the intergalactic medium.  The motion of this
intervening medium also introduces significant relative amplitude fluctuations (``scintillation'').  
Scintillation reprocesses both our target modulated signal and any colocated background.   
The amount of scintillation depends strongly on the line of sight to the source,  the plasma content and
motion near the merging binary, the observing frequency,   the angular size of the source, and the modulation timescale
of interest; see \cite{1998MNRAS.294..307W}, \cite{1990ARAA..28..561R} and references therein.  
Because of the wide range in possible parameters that are unrelated to the intrinsic nature of the source, we do not
model scintillation-induced backgrounds in detail  here.    

To some degree, modeling is unnecessary.  To the extent that background
AGN activity contaminates a signal, a phenomenological approach to
 AGN activity automatically includes all noise sources, scintillation included. 
Likewise, scintillation has characteristic power-law scalings of amplitude and timescale with
frequency \citep{1998MNRAS.294..307W}, allowing  identification and partial subtraction
of these modulations.

That said,  scintillation naturally produces $O(1)$ flux variations on an unknown but long timescale comparable to any periodicity
of interest.  For extended sources experiencing strong refractive scattering, 
scintillation naturally occurs on long timescales, 
roughly the time an intervening plasma element  $v$ would need to cross the source size $R$.  While the size of the
emission region is poorly known, the lower bound of the horizon size insures this crossing time is large 
\[t_{scin}\propto R/v
\simeq 20\unit{day} (M/10^8 M_\odot)(R/M)(50\unit{km/s}/v)
\]
where we scale $v$ to  typical galactic plasma velocities. 
In this case, typical scintillation flux variations should be large, with $\delta F/F\propto (\nu/\nu_o)^{17/30} \simeq O(1)$; based on  Figure 1
from \cite{1998MNRAS.294..307W} we expect a transition frequency  $\nu_o$  between 1 to ten times a fiducial
$1\unit{GHz}$ 
observing frequency, depending on the line of sight.
For  binary  candidates found by radio surveys, multiband followup observations  will be required to disentangle the impact of scintillation from
any intrinsic modulation.

\subsection{Unique time signatures versus confusing variability backgrounds}
As exemplified by OJ287 \citep{katz97}, an observed periodicity could be fit by other processes produced by  single black
holes, such as a warped disk causing the black hole to precess \citep{katz97}
or periodic accretion from or excited by a companion \citep[see,\,e.g.,\,\S~\ref{sec:context},][and references therein]{1996ApJ...467L..77A,2007PASJ...59..427H,2011PhRvD..84b4024F}.  Binary black holes can likewise produce periodic emission through other mechanisms,
including  orbiting accretion flows and modes excited in the circumbinary disk.
%
A detailed treatment of all possible backgrounds is beyond the scope of this paper.
However, the mere presence of an unmodeled periodicity alone does not uniquely determine this mechanism.

\begin{figure}
\includegraphics{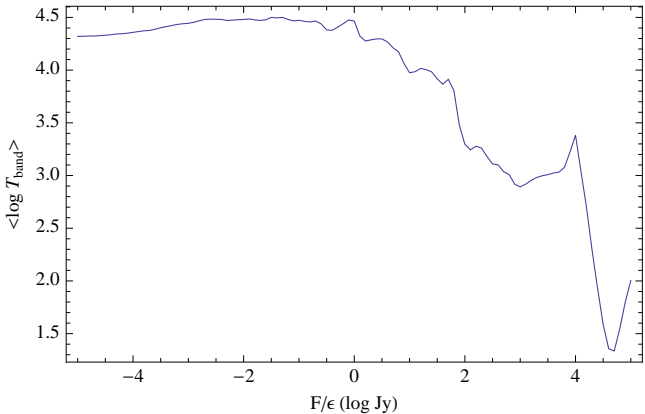}
\caption{\label{fig:LifetimeDiagram}\textbf{Typical lifetime}: The average log lifetime $\left<\log T\right>$ versus
  scaled flux $F/\epsilon$, including sources with observed orbital period $<1\unit{yr}$.  The brightest sources have
  short typical lifetimes, as they consist of the most massive nearby SMBH binaries in tight orbits.
}
\end{figure}

On the other hand, extremely sensitive surveys that recover many merging events may find a few with distinctive signatures: either (a) a
distinctive ultrarelativistic (i.e., Doppler-boosted and shapiro-delay-distorted) light curve; (b)
multiple periodic timescales;  or (c) a noticable chirp.     The most distinctive signature, a noticable chirp, occurs
surprisingly often among detectable binaries.   For example, an all-sky survey with $F/\epsilon=0.1\unit{Jy}$ can potentially observe between $300-1000$
binaries with less than $1\unit{yr}$ observed periods  [Fig. \ref{fig:EddingtonB:AllSkyBrightJets:OrbitPeriod}].  That population has a $\simeq 10\%$
probability of one member undergoing a chirp from $1\unit{yr}$ to $1\unit{min}$ within $10\unit{yr}$.
Moreover, the most massive and brightest members generally have shorter chirp times; see, e.g., the average log lifetime
in Figure \ref{fig:LifetimeDiagram}.

\subsection{Pulsar timing GW signature}
The massive, wide binaries which produce the most numerous and brightest jets are precisely the same sources whose
gravitational wave emission is targeted by pulsar timing arrays \citep{2009MNRAS.394.2255S,2010CQGra..27h4013H}.
Pulsar timing arrays are very likely to have a significantly shorter range than direct electromagnetic searches.
To order of magnitude, the ratio of radio to gravitational wave power is $\simeq M^2 r^4\omega^6/L_{edd}$, or
\begin{eqnarray}
\frac{L_{GW}}{L_j} &=& \frac{M \sigma_T}{m_p c^2}(2\pi/P)^2 \beta^2 
 \eqnbreak{\simeq} 3\times 10^6 \epsilon^{-1}(P/\unit{yr})^2 (M/10^8 M_\odot)  \beta^2
\end{eqnarray}
On the other hand, the relative energy flux sensitivity of a gravitational wave survey with characteristic strain sensitivity $h_s$
\begin{eqnarray}
\frac{F_{gw,min}}{F_{min}\nu} &=&  \frac{h_s^2 (2\pi/P)^2}{F_{em}\nu}
\end{eqnarray}
 As a result,
the relative luminosity distance sensitivity of pulsar timing arrays and EM surveys goes as
\begin{eqnarray}
\frac{d_{L,gw}}{d_{L,em}}  &=&\sqrt{\frac{F_{radio}\nu}{F_{GW}} \frac{L_{GW}}{L_{EM}}}  \\
 &=& h_s^{-1} \beta \sqrt{F_{radio}\nu M \sigma_T/m_p} \\
&\simeq& \frac{3\times 10^{-4}}{(h_s/10^{-13})} \beta  \sqrt{\frac{(F_{radio}/\unit{mJy}) (M/10^8 M_\odot) }{(\epsilon/0.002)}  }
\end{eqnarray}
where we have scaled $h_s$ to existing pulsar timing array sensitivities \citep{2010MNRAS.407..669Y}.  Future pulsar
timing arrays will have enough reach to identify sources up to their confusion limit.  Conservatively assuming only one source per
frequency can be distinguished \citep{2009MNRAS.394.2255S}, at least a few  sources over
the entire sky could be individually resolved.  Depending on how well the network localizes sources on the sky, the confusion limit
could be significantly lower \citep[cf.][]{2010arXiv1010.4337B}.
%
%
%
Nonetheless, barring extremely inefficient conversion of jet power to radio, direct electromagnetic surveys will have
substantially greater sensitivity than pulsar timing.
A pulsar timing survey by itself will only be more effective if the  efficiency $\epsilon$ and flux $F$ imply  $N(>F/\epsilon)$ is small -- in other words, only if radio
surveys can find at best a handful of sources  over the entire sky.

Nonetheless, electromagnetic and gravitational wave surveys naturally complement each other.
Owing to considerable uncertainties in the emission geometry and model, only a gravitational wave signature can confirm
a jet candidate corresponds to an SMBH binary.  
Conversely, starting with a well-localized binary candidate from pulsar timing \citep{2010PhRvD..81j4008S}, electromagnetic observations can
tightly localize a suitably-modulated companion \citep{gw-astro-pta-Sesana2011} with suitable overall circumbinary
spectra \citep{gw-astro-smbh-pta-Tanaka2011,2005ApJ...622L..93M}.
Subsequently, as observations provide a candidate light curve
corresponding to a known binary,
archival and directed EM surveys can find more, at higher confidence. 
Eventually, in the limit of perfect coupling, a joint EM and gravitational wave survey could construct joint confidence
in a proposed reconstruction of its data as a superposition of nearby and distant binaries plus background variability.

\section{Discussion}
Each SMBH binary should possess a pair of magnetically-powered bright jets attached to each hole during the last stages of
gravitational-wave driven inspiral.  Supermassive black holes therefore ubiquitously produce long-lived, modulated, and
moderately bright emission even in (local) vacuum, without direct accretion.  Though each jet is faint and though
mergers are rare, gravitational-wave-driven inspiral is so slow that many modulation cycles exist per binary.   Unlike
rare merger events  \citep{bbhpaper}, these modulations should easily be seen with future blind radio surveys like VAST
\citep{mc+11}, only excepting 
pessimistic choices for the efficiencies $\er$, $\eedd$ and obscuration $\fgeo$.
For example,  following \cite{2010Sci...329..927P} in adopting the conservative value $\eedd=0.002$, 
we predict 30-60 observable binaries with
observable orbital periods $<1\unit{yr}$ could be seen by the VAST survey ($10^4 \unit{deg}$ sky coverage, and either
1/day or 7/year cadence; see Figure \ref{fig:Modulation}).   Alternatively, if we include the diffuse emission found by
\citet{luciano-newjet}, then $\eedd$ is significantly larger, and roughly ten times more sources could be detected. 
Our results are insensitive to the precise merger rate we assume.

At any time, the most accessible sources are the largest black hole binaries in the widest ($\simeq $ year)  orbits.  Nonetheless, surveys with shorter
cadence can achieve comparable sensitivity by stacking cycles.  The gravitational
wave emission from these systems is also the target of pulsar timing arrays \citep{2009MNRAS.394.2255S}.
Unless very little jet power escapes in modulated radio jets,  radio surveys will generally be more
efficient than pulsar timing surveys for the same binaries.

The region surrounding a  SMBH offers many opportunities to produce variability on comparable timescales and
comparable bands.   In special but not exceptionally rare circumstances,  this mechanism
can be identified  by multiple timescales  -- depending on the scale, some combination of effects from its ultrarelativistic orbit (Doppler and Shapiro),
spin-orbit precession, and in exceptional circumstances the gravitational wave chirp.

If most  merging binaries accreted and produced long-lived AGN jets before decoupling from the circumbinary disk, then
this long-lived remnant likely has  significant variability which could  severely limit the effiacy of a radio survey.  
Targeted searches towards known jets must overcome similar variability challenges.

While we emphasize  and scale fiducial results to radio-band measurements; our expressions involve only flux sensitivity
limits and efficiencies and scale to any band.\footnote{At sufficiently high energies, less than one
photon arrives per orbit period in a reasonable-scale detector area.  This effect is significant only for extremely low
mass binaries in ultrarelativistic orbits, exceptionally high photon energies, or exceptionally low efficiencies: $M<0.5\times
10^3 M_\odot (d/\unit{Gpc}) (E_\gamma/\unit{keV})^{1/2} v^{3/2}/\epsilon^{1/2}$.  Even for $\epsilon\simeq 10^{-6}$, the most interesting
sources would still produce photon modulations in the X-ray.}
On the other hand, our preferred model of sinusoidal modulations visible equally well in all directions likely scales
to low-energy emission.   High energy emission naturally arises from an ultrarelativistic and tightly beamed outflow,
with dominant modulations from precession and  subdominant modulations from orbital motion (e.g., from 
curvature near the source, for marginally-detected lines of sight into tight orbits).  In that case, the probability of detecting prompt emission from any one
jet will be reduced by $\delta \theta^2/4\pi $.  On the other hand, the internal structure of ultrarelativistic jets
versus angle could magnify the effect of small changes in angle.  A handful of tightly beamed jets could exhibit a
distinctive rise and fall on the longest observing timescales. \hideindraft{ \editremark{Atish?}}

In this paper we only address the prospects for detecting a periodically modulated outflow in the radio, after the binary
decouples completely from its circumbinary disk.   We assume no accretion.  However, so long as gravitational radiation dominates the inspiral, our
calculations can be easily generalized to any partially- or totally- accretion-powered EM signal, just by  replacing
Eq. \ref{eq:def:JetPower} for $L_j(\beta)$ and by modeling the  modulation  $\delta(P)$ in Eq. \ref{eq:def:PeriodicDeltaSNR}.
For example, \citet{2010MNRAS.407.2007C} propose an inner circumbinary disk will be accreted onto a more massive
companion, with disk luminosity $L\propto T^{-5/4}\propto \beta^{10}$ peaking at $L\simeq 0.1 L_{\rm Edd}$.
Observations will easily distinguish between these generalizations: each generically predicts different scalings of
number versus flux, period, and modulation shape.  For example,    \citet{smbh-bin-HKM2009} discuss how the flux
distribution  $N(>F)$ could be used to extract information about the relative scaling of inspiral and emission with
orbit period. 

In our paper, we assume emission tracks the underlying SMBH dynamics; the outflow provides enough power for potentially
significant reach in a quiet band (e.g., $\epsilon>10^{-4}$); and 
 reprocessing or relativistic kinematics at most weakly distorts the signal.  
The details of the emission mechanism remain unknown and merit further detailed investigations.

For convenience, rather than discuss how the disk generates a magnetic field and how the outflow produces prompt radio
power, we have adopted a universal efficiency $\epsilon=\eedd \er$  for all binaries, regardless of component masses,
separation, or disk structure.   Each product in the composite efficiency likely depends on binary and disk properties.
For example, the efficiency $\eedd$ is equivalent to the strength of the field threading the horizon; as described in Footnote \ref{foot:Main}, this field strength depends on the disk structure and accretion flow.
Likewise, the radio efficiency $\er$ depends on the emission process, in turn possibly  dependent on the amount of
coronal matter in the interior.
Further investigations are needed to quantify the magnitude, variability, and parameter dependence of this composite efficiency.

Finally, we have implicitly made strong assumptions about magnetic field order, particularly for the widest unequal-mass
ratio binaries.  The passage of the black hole through the magnetic field liberates the local
instantaneous energy density, possibly beamed along the instantaneous ambient field direction.   So long as the magnetic
field is steady in magnitude and direction during all observations $T_{obs}$,  only the orbit's kinematics and outflow
variability lead to
modulations in the observed flux.  However, for the widest orbits close to the circumbinary disk, each black  hole will
move through a spatially- and temporally- fluctuating field, causing additional variability.  
Detectable sources with such wide orbits should be rare [Figure \ref{fig:TimescaleDiagram}].  
Future  MHD simulations of  circumbinary disks should extract the correlation functions of magnetic
flux in the circumbinary region, to better determine the level of variability expected.

\begin{acknowledgements}
DLK and AK are partially supported by NSF award AST-1008353. ROS is supported by NSF award PHY-0970074, the Bradley
Program Fellowship (BCS),  and the UWM Research
Growth Initiative.  
AS is supported by the Max Planck Society.
The authors thank M. Eracleous, J. Krolik, H. Bignall, L. Rezzolla, P. Moesta, D. Alic, O. Zanotti, and J. Lazio  for helpful feedback, as well as the
anonymous referee for thoughtful suggestions.  ROS and AS also thank Aspen Center for Physics, where
this work was completed, and all participants of its  2011  program on Galaxy and Supermassive Black Hole Coevolution.
\end{acknowledgements}

\bibliographystyle{astroads}  
\bibliography{prec}

\begin{thebibliography}{75}
\expandafter\ifx\csname natexlab\endcsname\relax\def\natexlab#1{#1}\fi
\expandafter\ifx\csname href\endcsname\relax
  \def\href#1#2{}\fi
\expandafter\ifx\csname urllinklabel\endcsname\relax
  \def\urllinklabel{[URL]}\fi
\expandafter\ifx\csname adsurllinklabel\endcsname\relax
  \def\adsurllinklabel{[ADS]}\fi

\bibitem[{{Apostolatos} {et~al.}(1994){Apostolatos}, {Cutler}, {Sussman}, \&
  {Thorne}}]{acst94}
{Apostolatos}, T.~A., {Cutler}, C., {Sussman}, G.~J., \& {Thorne}, K.~S. 1994,
  \prd, 49, 6274

 \href{http://adsabs.harvard.edu/abs/1994PhRvD..49.6274A}{\adsurllinklabel}

\bibitem[{{Artymowicz} \& {Lubow}(1996)}]{1996ApJ...467L..77A}
{Artymowicz}, P. \& {Lubow}, S.~H. 1996, \apjl, 467, L77+

 \href{http://adsabs.harvard.edu/abs/1996ApJ...467L..77A}{\adsurllinklabel}

\bibitem[{{Arun} {et~al.}(2009){Arun}, {Babak}, {Berti}, {Cornish}, {Cutler},
  {Gair}, {Hughes}, {Iyer}, {Lang}, {Mandel}, {Porter}, {Sathyaprakash},
  {Sinha}, {Sintes}, {Trias}, {Van Den Broeck}, \& {Volonteri}}]{abb+09}
{Arun}, K.~G., {Babak}, S., {Berti}, E., {Cornish}, N., {Cutler}, C., {Gair},
  J., {Hughes}, S.~A., {Iyer}, B.~R., {Lang}, R.~N., {Mandel}, I., {Porter},
  E.~K., {Sathyaprakash}, B.~S., {Sinha}, S., {Sintes}, A.~M., {Trias}, M.,
  {Van Den Broeck}, C., \& {Volonteri}, M. 2009, Classical and Quantum Gravity,
  26, 094027

 \href{http://adsabs.harvard.edu/abs/2009CQGra..26i4027A}{\adsurllinklabel}

\bibitem[{{Beckwith} {et~al.}(2009){Beckwith}, {Hawley}, \&
  {Krolik}}]{2009ApJ...707..428B}
{Beckwith}, K., {Hawley}, J.~F., \& {Krolik}, J.~H. 2009, \apj, 707, 428

 \href{http://adsabs.harvard.edu/abs/2009ApJ...707..428B}{\adsurllinklabel}

\bibitem[{{Bertone} \& {Conselice}(2009)}]{2009MNRAS.396.2345B}
{Bertone}, S. \& {Conselice}, C.~J. 2009, \mnras, 396, 2345

 \href{http://adsabs.harvard.edu/abs/2009MNRAS.396.2345B}{\adsurllinklabel}

\bibitem[{{Bloom} \& {et al}(2009)}]{whitepaper-CoordinatedScience}
{Bloom}, J.~S. \& {et al}. 2009, arXiv:0902.1527

 \href{http://adsabs.harvard.edu/abs/2009arXiv0902.1527B}{\adsurllinklabel}

\bibitem[{{Bode} {et~al.}(2011){Bode}, {Bogdanovic}, {Haas}, {Healy}, {Laguna},
  \& {Shoemaker}}]{smbh-bin-merger-EMC-Disk-BodeBogdanovic-2011}
{Bode}, T., {Bogdanovic}, T., {Haas}, R., {Healy}, J., {Laguna}, P., \&
  {Shoemaker}, D. 2011, arXiv:1101.4684
 \href{http://xxx.lanl.gov/abs/arXiv:1101.4684}{\urllinklabel}


\bibitem[{{Bogdanovi{\'c}} {et~al.}(2009){Bogdanovi{\'c}}, {Eracleous}, \&
  {Sigurdsson}}]{2009ApJ...697..288B}
{Bogdanovi{\'c}}, T., {Eracleous}, M., \& {Sigurdsson}, S. 2009, \apj, 697, 288

 \href{http://adsabs.harvard.edu/abs/2009ApJ...697..288B}{\adsurllinklabel}

\bibitem[{{Bogdanovi{\'c}} {et~al.}(2007){Bogdanovi{\'c}}, {Reynolds}, \&
  {Miller}}]{2007ApJ...661L.147B}
{Bogdanovi{\'c}}, T., {Reynolds}, C.~S., \& {Miller}, M.~C. 2007, \apjl, 661,
  L147

 \href{http://adsabs.harvard.edu/abs/2007ApJ...661L.147B}{\adsurllinklabel}

\bibitem[{{Boroson} \& {Lauer}(2009)}]{2009Natur.458...53B}
{Boroson}, T.~A. \& {Lauer}, T.~R. 2009, \nat, 458, 53

 \href{http://adsabs.harvard.edu/abs/2009Natur.458...53B}{\adsurllinklabel}

\bibitem[{{Boyle} \& {Pen}(2010)}]{2010arXiv1010.4337B}
{Boyle}, L. \& {Pen}, U.-L. 2010, ArXiv e-prints

 \href{http://adsabs.harvard.edu/abs/2010arXiv1010.4337B}{\adsurllinklabel}

\bibitem[{{Burke-Spolaor}(2010)}]{bs10}
{Burke-Spolaor}, S. 2010, \mnras, 1574

 \href{http://adsabs.harvard.edu/abs/2010MNRAS.tmp.1574B}{\adsurllinklabel}

\bibitem[{{Chang} {et~al.}(2010){Chang}, {Strubbe}, {Menou}, \&
  {Quataert}}]{2010MNRAS.407.2007C}
{Chang}, P., {Strubbe}, L.~E., {Menou}, K., \& {Quataert}, E. 2010, \mnras,
  407, 2007

 \href{http://adsabs.harvard.edu/abs/2010MNRAS.407.2007C}{\adsurllinklabel}

\bibitem[{{Chen} {et~al.}(2009){Chen}, {Madau}, {Sesana}, \&
  {Liu}}]{2009ApJ...697L.149C}
{Chen}, X., {Madau}, P., {Sesana}, A., \& {Liu}, F.~K. 2009, \apjl, 697, L149

 \href{http://adsabs.harvard.edu/abs/2009ApJ...697L.149C}{\adsurllinklabel}

\bibitem[{{Comerford} {et~al.}(2009){Comerford}, {Gerke}, {Newman}, {Davis},
  {Yan}, {Cooper}, {Faber}, {Koo}, {Coil}, {Rosario}, \&
  {Dutton}}]{2009ApJ...698..956C}
{Comerford}, J.~M., {Gerke}, B.~F., {Newman}, J.~A., {Davis}, M., {Yan}, R.,
  {Cooper}, M.~C., {Faber}, S.~M., {Koo}, D.~C., {Coil}, A.~L., {Rosario},
  D.~J., \& {Dutton}, A.~A. 2009, \apj, 698, 956

 \href{http://adsabs.harvard.edu/abs/2009ApJ...698..956C}{\adsurllinklabel}

\bibitem[{{Demorest} {et~al.}(2009){Demorest}, {Lazio}, {Lommen}, {Archibald},
  {Arzoumanian}, {Backer}, {Cordes}, {Demorest}, {Ferdman}, {Freire},
  {Gonzalez}, {Jenet}, {Kaspi}, {Kondratiev}, {Lazio}, {Lommen}, {Lorimer},
  {Lynch}, {McLaughlin}, {Nice}, {Ransom}, {Shannon}, {Siemens}, {Stairs},
  {Stinebring}, {Reitze}, {Shoemaker}, {Whitcomb}, \&
  {Weiss}}]{whitepaper-PulsarTiming2}
{Demorest}, P., {Lazio}, J., {Lommen}, A., {Archibald}, A., {Arzoumanian}, Z.,
  {Backer}, D., {Cordes}, J., {Demorest}, P., {Ferdman}, R., {Freire}, P.,
  {Gonzalez}, M., {Jenet}, R., {Kaspi}, V., {Kondratiev}, V., {Lazio}, J.,
  {Lommen}, A., {Lorimer}, D., {Lynch}, R., {McLaughlin}, M., {Nice}, D.,
  {Ransom}, S., {Shannon}, R., {Siemens}, X., {Stairs}, I., {Stinebring}, D.,
  {Reitze}, D., {Shoemaker}, D., {Whitcomb}, S., \& {Weiss}, R. 2009, in
  Astronomy, Vol. 2010, AGB Stars and Related Phenomenastro2010: The Astronomy
  and Astrophysics Decadal Survey, 64--+


\bibitem[{{Dotti} {et~al.}(2009){Dotti}, {Montuori}, {Decarli}, {Volonteri},
  {Colpi}, \& {Haardt}}]{2009MNRAS.398L..73D}
{Dotti}, M., {Montuori}, C., {Decarli}, R., {Volonteri}, M., {Colpi}, M., \&
  {Haardt}, F. 2009, \mnras, 398, L73

 \href{http://adsabs.harvard.edu/abs/2009MNRAS.398L..73D}{\adsurllinklabel}

\bibitem[{{Eracleous} {et~al.}(2011){Eracleous}, {Boroson}, {Halpern}, \&
  {Liu}}]{smbh-Eracleous-SurveyForClose-2011}
{Eracleous}, M., {Boroson}, T.~A., {Halpern}, J.~P., \& {Liu}, J. 2011,
  (arXiv:1106.2952)
 \href{http://xxx.lanl.gov/abs/arXiv:1106.2952}{\urllinklabel}


\bibitem[{{Farris} {et~al.}(2011){Farris}, {Liu}, \&
  {Shapiro}}]{2011PhRvD..84b4024F}
{Farris}, B.~D., {Liu}, Y.~T., \& {Shapiro}, S.~L. 2011, \prd, 84, 024024

 \href{http://adsabs.harvard.edu/abs/2011PhRvD..84b4024F}{\adsurllinklabel}

\bibitem[{{Haiman} {et~al.}(2009{\natexlab{a}}){Haiman}, {Kocsis}, \&
  {Menou}}]{smbh-bin-HKM2009}
{Haiman}, Z., {Kocsis}, B., \& {Menou}, K. 2009{\natexlab{a}},
  (arXiv:0904.1383)
 \href{http://xxx.lanl.gov/abs/arXiv:0904.1383}{\urllinklabel}


\bibitem[{{Haiman} {et~al.}(2009{\natexlab{b}}){Haiman}, {Kocsis}, {Menou},
  {Lippai}, \& {Frei}}]{2009CQGra..26i4032H}
{Haiman}, Z., {Kocsis}, B., {Menou}, K., {Lippai}, Z., \& {Frei}, Z.
  2009{\natexlab{b}}, Classical and Quantum Gravity, 26, 094032

 \href{http://adsabs.harvard.edu/abs/2009CQGra..26i4032H}{\adsurllinklabel}

\bibitem[{{Hayasaki} {et~al.}(2007){Hayasaki}, {Mineshige}, \&
  {Sudou}}]{2007PASJ...59..427H}
{Hayasaki}, K., {Mineshige}, S., \& {Sudou}, H. 2007, \pasj, 59, 427

 \href{http://adsabs.harvard.edu/abs/2007PASJ...59..427H}{\adsurllinklabel}

\bibitem[{{Hobbs} {et~al.}(2010){Hobbs}, {Archibald}, {Arzoumanian}, {Backer},
  {Bailes}, {Bhat}, {Burgay}, {Burke-Spolaor}, {Champion}, {Cognard}, {Coles},
  {Cordes}, {Demorest}, {Desvignes}, {Ferdman}, {Finn}, {Freire}, {Gonzalez},
  {Hessels}, {Hotan}, {Janssen}, {Jenet}, {Jessner}, {Jordan}, {Kaspi},
  {Kramer}, {Kondratiev}, {Lazio}, {Lazaridis}, {Lee}, {Levin}, {Lommen},
  {Lorimer}, {Lynch}, {Lyne}, {Manchester}, {McLaughlin}, {Nice}, {Oslowski},
  {Pilia}, {Possenti}, {Purver}, {Ransom}, {Reynolds}, {Sanidas}, {Sarkissian},
  {Sesana}, {Shannon}, {Siemens}, {Stairs}, {Stappers}, {Stinebring},
  {Theureau}, {van Haasteren}, {van Straten}, {Verbiest}, {Yardley}, \&
  {You}}]{2010CQGra..27h4013H}
{Hobbs}, G., {Archibald}, A., {Arzoumanian}, Z., {Backer}, D., {Bailes}, M.,
  {Bhat}, N.~D.~R., {Burgay}, M., {Burke-Spolaor}, S., {Champion}, D.,
  {Cognard}, I., {Coles}, W., {Cordes}, J., {Demorest}, P., {Desvignes}, G.,
  {Ferdman}, R.~D., {Finn}, L., {Freire}, P., {Gonzalez}, M., {Hessels}, J.,
  {Hotan}, A., {Janssen}, G., {Jenet}, F., {Jessner}, A., {Jordan}, C.,
  {Kaspi}, V., {Kramer}, M., {Kondratiev}, V., {Lazio}, J., {Lazaridis}, K.,
  {Lee}, K.~J., {Levin}, Y., {Lommen}, A., {Lorimer}, D., {Lynch}, R., {Lyne},
  A., {Manchester}, R., {McLaughlin}, M., {Nice}, D., {Oslowski}, S., {Pilia},
  M., {Possenti}, A., {Purver}, M., {Ransom}, S., {Reynolds}, J., {Sanidas},
  S., {Sarkissian}, J., {Sesana}, A., {Shannon}, R., {Siemens}, X., {Stairs},
  I., {Stappers}, B., {Stinebring}, D., {Theureau}, G., {van Haasteren}, R.,
  {van Straten}, W., {Verbiest}, J.~P.~W., {Yardley}, D.~R.~B., \& {You}, X.~P.
  2010, Classical and Quantum Gravity, 27, 084013

 \href{http://adsabs.harvard.edu/abs/2010CQGra..27h4013H}{\adsurllinklabel}

\bibitem[{{Hughes} {et~al.}(1992){Hughes}, {Aller}, \&
  {Aller}}]{1992ApJ...396..469H}
{Hughes}, P.~A., {Aller}, H.~D., \& {Aller}, M.~F. 1992, \apj, 396, 469

 \href{http://adsabs.harvard.edu/abs/1992ApJ...396..469H}{\adsurllinklabel}

\bibitem[{{Hughes}(2002)}]{2002MNRAS.331..805H}
{Hughes}, S.~A. 2002, \mnras, 331, 805

 \href{http://adsabs.harvard.edu/abs/2002MNRAS.331..805H}{\adsurllinklabel}

\bibitem[{{Jenet} {et~al.}(2009){Jenet}, {Finn}, {Lazio}, {Lommen},
  {McLaughlin}, {Stairs}, {Stinebring}, {Verbiest}, {Archibald}, {Arzoumanian},
  {Backer}, {Cordes}, {Demorest}, {Ferdman}, {Freire}, {Gonzalez}, {Kaspi},
  {Kondratiev}, {Lorimer}, {Lynch}, {Nice}, {Ransom}, {Shannon}, \&
  {Siemens}}]{whitepaper-PulsarTiming1}
{Jenet}, F., {Finn}, L.~S., {Lazio}, J., {Lommen}, A., {McLaughlin}, M.,
  {Stairs}, I., {Stinebring}, D., {Verbiest}, J., {Archibald}, A.,
  {Arzoumanian}, Z., {Backer}, D., {Cordes}, J., {Demorest}, P., {Ferdman}, R.,
  {Freire}, P., {Gonzalez}, M., {Kaspi}, V., {Kondratiev}, V., {Lorimer}, D.,
  {Lynch}, R., {Nice}, D., {Ransom}, S., {Shannon}, R., \& {Siemens}, X. 2009,
  ArXiv e-prints

 \href{http://adsabs.harvard.edu/abs/2009arXiv0909.1058J}{\adsurllinklabel}

\bibitem[{{Johnston} {et~al.}(2007)}]{jbb+07}
{Johnston}, S. {et~al.} 2007, \pasa, 24, 174

 \href{http://adsabs.harvard.edu/abs/2007PASA...24..174J}{\adsurllinklabel}

\bibitem[{{Kaplan} {et~al.}(2011){Kaplan}, {O'Shaughnessy}, {Sesana}, \&
  {Volonteri}}]{bbhpaper}
{Kaplan}, D.~L., {O'Shaughnessy}, R., {Sesana}, A., \& {Volonteri}, M. 2011,
  \apjl, 734, L37+

 \href{http://adsabs.harvard.edu/abs/2011ApJ...734L..37K}{\adsurllinklabel}

\bibitem[{{Katz}(1997)}]{katz97}
{Katz}, J.~I. 1997, \apj, 478, 527

 \href{http://adsabs.harvard.edu/abs/1997ApJ...478..527K}{\adsurllinklabel}

\bibitem[{{King} {et~al.}(2007){King}, {Pringle}, \&
  {Livio}}]{2007MNRAS.376.1740K}
{King}, A.~R., {Pringle}, J.~E., \& {Livio}, M. 2007, \mnras, 376, 1740

 \href{http://adsabs.harvard.edu/abs/2007MNRAS.376.1740K}{\adsurllinklabel}

\bibitem[{{Komossa}(2006)}]{komossa06}
{Komossa}, S. 2006, \memsai, 77, 733

 \href{http://adsabs.harvard.edu/abs/2006MmSAI..77..733K}{\adsurllinklabel}

\bibitem[{{Komossa} {et~al.}(2003){Komossa}, {Burwitz}, {Hasinger}, {Predehl},
  {Kaastra}, \& {Ikebe}}]{2003ApJ...582L..15K}
{Komossa}, S., {Burwitz}, V., {Hasinger}, G., {Predehl}, P., {Kaastra}, J.~S.,
  \& {Ikebe}, Y. 2003, \apjl, 582, L15

 \href{http://adsabs.harvard.edu/abs/2003ApJ...582L..15K}{\adsurllinklabel}

\bibitem[{Krolik(1998)}]{book-Krolik-AGN}
Krolik, J. 1998, Active Galactic Nuclei (Princeton Series in Astrophysics)


\bibitem[{{Lawrence} \& {Elvis}(2010)}]{2010ApJ...714..561L}
{Lawrence}, A. \& {Elvis}, M. 2010, \apj, 714, 561

 \href{http://adsabs.harvard.edu/abs/2010ApJ...714..561L}{\adsurllinklabel}

\bibitem[{{Loeb} \& {Gaudi}(2003)}]{2003ApJ...588L.117L}
{Loeb}, A. \& {Gaudi}, B.~S. 2003, \apjl, 588, L117

 \href{http://adsabs.harvard.edu/abs/2003ApJ...588L.117L}{\adsurllinklabel}

\bibitem[{{Lyutikov}(2011)}]{2011PhRvD..83f4001L}
{Lyutikov}, M. 2011, \prd, 83, 064001

 \href{http://adsabs.harvard.edu/abs/2011PhRvD..83f4001L}{\adsurllinklabel}

\bibitem[{{Marscher} \& {Jorstad}(2011)}]{2011ApJ...729...26M}
{Marscher}, A.~P. \& {Jorstad}, S.~G. 2011, \apj, 729, 26

 \href{http://adsabs.harvard.edu/abs/2011ApJ...729...26M}{\adsurllinklabel}

\bibitem[{{Megevand} {et~al.}(2009){Megevand}, {Anderson}, {Frank},
  {Hirschmann}, {Lehner}, {Liebling}, {Motl}, \&
  {Neilsen}}]{2009PhRvD..80b4012M}
{Megevand}, M., {Anderson}, M., {Frank}, J., {Hirschmann}, E.~W., {Lehner}, L.,
  {Liebling}, S.~L., {Motl}, P.~M., \& {Neilsen}, D. 2009, \prd, 80, 024012

 \href{http://adsabs.harvard.edu/abs/2009PhRvD..80b4012M}{\adsurllinklabel}

\bibitem[{{Milosavljevi{\'c}} \& {Phinney}(2005)}]{2005ApJ...622L..93M}
{Milosavljevi{\'c}}, M. \& {Phinney}, E.~S. 2005, \apjl, 622, L93

 \href{http://adsabs.harvard.edu/abs/2005ApJ...622L..93M}{\adsurllinklabel}

\bibitem[{{Moesta} {et~al.}(2011){Moesta}, {Alic}, { Rezzolla}, \&
  {Zanotti}}]{luciano-newjet}
{Moesta}, P., {Alic}, D., { Rezzolla}, L., \& {Zanotti}, O. 2011, arXiv
  preprints



\bibitem[{{Murphy} {et~al.}(2011){Murphy}, {Chatterjee}, {et~al.}}]{mc+11}
{Murphy}, T., {Chatterjee}, S., {et~al.} 2011, \pasa, in prep.



\bibitem[{{Ofek} {et~al.}(2011){Ofek}, {Frail}, {Breslauer}, {Kulkarni},
  {Chandra}, {Gal-Yam}, {Kasliwal}, \& {Gehrels}}]{ofb+11}
{Ofek}, E.~O., {Frail}, D.~A., {Breslauer}, B., {Kulkarni}, S.~R., {Chandra},
  P., {Gal-Yam}, A., {Kasliwal}, M.~M., \& {Gehrels}, N. 2011, \apj,
  arXiv:1103.3010

 \href{http://adsabs.harvard.edu/abs/2011arXiv1103.3010O}{\adsurllinklabel}

\bibitem[{{O'Neill} {et~al.}(2009){O'Neill}, {Miller}, {Bogdanovi{\'c}},
  {Reynolds}, \& {Schnittman}}]{2009ApJ...700..859O}
{O'Neill}, S.~M., {Miller}, M.~C., {Bogdanovi{\'c}}, T., {Reynolds}, C.~S., \&
  {Schnittman}, J.~D. 2009, \apj, 700, 859

 \href{http://adsabs.harvard.edu/abs/2009ApJ...700..859O}{\adsurllinklabel}

\bibitem[{{Palenzuela} {et~al.}(2010{\natexlab{a}}){Palenzuela}, {Lehner}, \&
  {Liebling}}]{pll10}
{Palenzuela}, C., {Lehner}, L., \& {Liebling}, S.~L. 2010{\natexlab{a}},
  Science, 329, 927

 \href{http://adsabs.harvard.edu/abs/2010Sci...329..927P}{\adsurllinklabel}

\bibitem[{{Palenzuela} {et~al.}(2010{\natexlab{b}}){Palenzuela}, {Lehner}, \&
  {Liebling}}]{2010Sci...329..927P}
---. 2010{\natexlab{b}}, Science, 329, 927

 \href{http://adsabs.harvard.edu/abs/2010Sci...329..927P}{\adsurllinklabel}

\bibitem[{{Pessah} {et~al.}(2006){Pessah}, {Chan}, \&
  {Psaltis}}]{2006PhRvL..97v1103P}
{Pessah}, M.~E., {Chan}, C., \& {Psaltis}, D. 2006, Physical Review Letters,
  97, 221103

 \href{http://adsabs.harvard.edu/abs/2006PhRvL..97v1103P}{\adsurllinklabel}

\bibitem[{{Rickett}(1990)}]{1990ARAA..28..561R}
{Rickett}, B.~J. 1990, \araa, 28, 561

 \href{http://adsabs.harvard.edu/abs/1990ARA%26A..28..561R}{\adsurllinklabel}

\bibitem[{{Rodriguez} {et~al.}(2006{\natexlab{a}}){Rodriguez}, {Taylor},
  {Zavala}, {Peck}, {Pollack}, \& {Romani}}]{rtz+06}
{Rodriguez}, C., {Taylor}, G.~B., {Zavala}, R.~T., {Peck}, A.~B., {Pollack},
  L.~K., \& {Romani}, R.~W. 2006{\natexlab{a}}, \apj, 646, 49

 \href{http://adsabs.harvard.edu/abs/2006ApJ...646...49R}{\adsurllinklabel}

\bibitem[{{Rodriguez} {et~al.}(2006{\natexlab{b}}){Rodriguez}, {Taylor},
  {Zavala}, {Peck}, {Pollack}, \& {Romani}}]{2006ApJ...646...49R}
---. 2006{\natexlab{b}}, \apj, 646, 49

 \href{http://adsabs.harvard.edu/abs/2006ApJ...646...49R}{\adsurllinklabel}

\bibitem[{{Roedig} {et~al.}(2011){Roedig}, {Dotti}, {Sesana}, {Cuadra}, \&
  {Colpi}}]{2011MNRAS-Roedig}
{Roedig}, C., {Dotti}, M., {Sesana}, A., {Cuadra}, J., \& {Colpi}, M. 2011,
  \mnras, 979

 \href{http://adsabs.harvard.edu/abs/2011MNRAS.tmp..979R}{\adsurllinklabel}

\bibitem[{{Rossi} {et~al.}(2010){Rossi}, {Lodato}, {Armitage}, {Pringle}, \&
  {King}}]{2010MNRAS.401.2021R}
{Rossi}, E.~M., {Lodato}, G., {Armitage}, P.~J., {Pringle}, J.~E., \& {King},
  A.~R. 2010, \mnras, 401, 2021

 \href{http://adsabs.harvard.edu/abs/2010MNRAS.401.2021R}{\adsurllinklabel}

\bibitem[{{Rothstein} \& {Lovelace}(2008)}]{2008ApJ...677.1221R}
{Rothstein}, D.~M. \& {Lovelace}, R.~V.~E. 2008, \apj, 677, 1221

 \href{http://adsabs.harvard.edu/abs/2008ApJ...677.1221R}{\adsurllinklabel}

\bibitem[{{Schnittman}(2010)}]{review-Schnittman2010}
{Schnittman}, J.~D. 2010, ArXiv e-prints

 \href{http://adsabs.harvard.edu/abs/2010arXiv1010.3250S}{\adsurllinklabel}

\bibitem[{{Schnittman} \& {Krolik}(2008)}]{2008ApJ...684..835S}
{Schnittman}, J.~D. \& {Krolik}, J.~H. 2008, \apj, 684, 835
 \href{http://xxx.lanl.gov/abs/arXiv:0802.3556}{\urllinklabel}


\bibitem[{{Sesana} {et~al.}(2011{\natexlab{a}}){Sesana}, {Gair}, {Berti}, \&
  {Volonteri}}]{2011PhRvD..83d4036S}
{Sesana}, A., {Gair}, J., {Berti}, E., \& {Volonteri}, M. 2011{\natexlab{a}},
  \prd, 83, 044036

 \href{http://adsabs.harvard.edu/abs/2011PhRvD..83d4036S}{\adsurllinklabel}

\bibitem[{{Sesana} {et~al.}(2011{\natexlab{b}}){Sesana}, {Roedig}, {Reynolds},
  \& {Dotti}}]{gw-astro-pta-Sesana2011}
{Sesana}, A., {Roedig}, C., {Reynolds}, M.~T., \& {Dotti}, M.
  2011{\natexlab{b}}, ArXiv e-prints

 \href{http://adsabs.harvard.edu/abs/2011arXiv1107.2927S}{\adsurllinklabel}

\bibitem[{{Sesana} \& {Vecchio}(2010)}]{2010PhRvD..81j4008S}
{Sesana}, A. \& {Vecchio}, A. 2010, \prd, 81, 104008

 \href{http://adsabs.harvard.edu/abs/2010PhRvD..81j4008S}{\adsurllinklabel}

\bibitem[{{Sesana} {et~al.}(2009){Sesana}, {Vecchio}, \&
  {Volonteri}}]{2009MNRAS.394.2255S}
{Sesana}, A., {Vecchio}, A., \& {Volonteri}, M. 2009, \mnras, 394, 2255

 \href{http://adsabs.harvard.edu/abs/2009MNRAS.394.2255S}{\adsurllinklabel}

\bibitem[{{Sesana} {et~al.}(2007){Sesana}, {Volonteri}, \&
  {Haardt}}]{2007MNRAS.377.1711S}
{Sesana}, A., {Volonteri}, M., \& {Haardt}, F. 2007, \mnras, 377, 1711

 \href{http://adsabs.harvard.edu/abs/2007MNRAS.377.1711S}{\adsurllinklabel}

\bibitem[{{Shen} \& {Loeb}(2010)}]{2010ApJ...725..249S}
{Shen}, Y. \& {Loeb}, A. 2010, \apj, 725, 249

 \href{http://adsabs.harvard.edu/abs/2010ApJ...725..249S}{\adsurllinklabel}

\bibitem[{{Sillanpaa} {et~al.}(1988){Sillanpaa}, {Haarala}, {Valtonen},
  {Sundelius}, \& {Byrd}}]{1988ApJ...325..628S}
{Sillanpaa}, A., {Haarala}, S., {Valtonen}, M.~J., {Sundelius}, B., \& {Byrd},
  G.~G. 1988, \apj, 325, 628

 \href{http://adsabs.harvard.edu/abs/1988ApJ...325..628S}{\adsurllinklabel}

\bibitem[{{Smith} {et~al.}(2010){Smith}, {Shields}, {Bonning}, {McMullen},
  {Rosario}, \& {Salviander}}]{ssb+10}
{Smith}, K.~L., {Shields}, G.~A., {Bonning}, E.~W., {McMullen}, C.~C.,
  {Rosario}, D.~J., \& {Salviander}, S. 2010, \apj, 716, 866

 \href{http://adsabs.harvard.edu/abs/2010ApJ...716..866S}{\adsurllinklabel}

\bibitem[{{Stone} \&
  {Loeb}(2011)}]{clus-smbh-TidalDisruptionRates-StoneLoeb2011}
{Stone}, N. \& {Loeb}, A. 2011, \mnras, 224

 \href{http://adsabs.harvard.edu/abs/2011MNRAS.tmp..224S}{\adsurllinklabel}

\bibitem[{{Tanaka} \& {Menou}(2010)}]{2010ApJ...714..404T}
{Tanaka}, T. \& {Menou}, K. 2010, \apj, 714, 404

 \href{http://adsabs.harvard.edu/abs/2010ApJ...714..404T}{\adsurllinklabel}

\bibitem[{{Tanaka} {et~al.}(2011){Tanaka}, {Menou}, \&
  {Haiman}}]{gw-astro-smbh-pta-Tanaka2011}
{Tanaka}, T., {Menou}, K., \& {Haiman}, Z. 2011, ArXiv e-prints

 \href{http://adsabs.harvard.edu/abs/2011arXiv1107.2937T}{\adsurllinklabel}

\bibitem[{{Tsalmantza} {et~al.}(2011){Tsalmantza}, {Decarli}, {Dotti}, \&
  {Hogg}}]{2011arXiv1106.1180T}
{Tsalmantza}, P., {Decarli}, R., {Dotti}, M., \& {Hogg}, D.~W. 2011, ArXiv
  e-prints

 \href{http://adsabs.harvard.edu/abs/2011arXiv1106.1180T}{\adsurllinklabel}

\bibitem[{{Urry} \& {Padovani}(1995)}]{1995PASP..107..803U}
{Urry}, C.~M. \& {Padovani}, P. 1995, \pasp, 107, 803

 \href{http://adsabs.harvard.edu/abs/1995PASP..107..803U}{\adsurllinklabel}

\bibitem[{{Valtonen} {et~al.}(2008){Valtonen}, {Lehto}, {Nilsson}, {Heidt},
  {Takalo}, {Sillanp{\"a}{\"a}}, {Villforth}, {Kidger}, {Poyner}, {Pursimo},
  {Zola}, {Wu}, {Zhou}, {Sadakane}, {Drozdz}, {Koziel}, {Marchev}, {Ogloza},
  {Porowski}, {Siwak}, {Stachowski}, {Winiarski}, {Hentunen}, {Nissinen},
  {Liakos}, \& {Dogru}}]{2008Natur.452..851V}
{Valtonen}, M.~J., {Lehto}, H.~J., {Nilsson}, K., {Heidt}, J., {Takalo}, L.~O.,
  {Sillanp{\"a}{\"a}}, A., {Villforth}, C., {Kidger}, M., {Poyner}, G.,
  {Pursimo}, T., {Zola}, S., {Wu}, J.-H., {Zhou}, X., {Sadakane}, K., {Drozdz},
  M., {Koziel}, D., {Marchev}, D., {Ogloza}, W., {Porowski}, C., {Siwak}, M.,
  {Stachowski}, G., {Winiarski}, M., {Hentunen}, V.-P., {Nissinen}, M.,
  {Liakos}, A., \& {Dogru}, S. 2008, \nat, 452, 851

 \href{http://adsabs.harvard.edu/abs/2008Natur.452..851V}{\adsurllinklabel}

\bibitem[{{Volonteri}(2010)}]{2010AARv..18..279V}
{Volonteri}, M. 2010, \aapr, 18, 279



\bibitem[{{Volonteri} \& {Begelman}(2010)}]{2010MNRAS.409.1022V}
{Volonteri}, M. \& {Begelman}, M.~C. 2010, \mnras, 409, 1022

 \href{http://adsabs.harvard.edu/abs/2010MNRAS.409.1022V}{\adsurllinklabel}

\bibitem[{{Volonteri} {et~al.}(2003){Volonteri}, {Haardt}, \&
  {Madau}}]{2003ApJ...582..559V}
{Volonteri}, M., {Haardt}, F., \& {Madau}, P. 2003, \apj, 582, 559

 \href{http://adsabs.harvard.edu/abs/2003ApJ...582..559V}{\adsurllinklabel}

\bibitem[{{Volonteri} {et~al.}(2008){Volonteri}, {Lodato}, \&
  {Natarajan}}]{2008MNRAS.383.1079V}
{Volonteri}, M., {Lodato}, G., \& {Natarajan}, P. 2008, \mnras, 383, 1079

 \href{http://adsabs.harvard.edu/abs/2008MNRAS.383.1079V}{\adsurllinklabel}

\bibitem[{{Walker}(1998)}]{1998MNRAS.294..307W}
{Walker}, M.~A. 1998, \mnras, 294, 307

 \href{http://adsabs.harvard.edu/abs/1998MNRAS.294..307W}{\adsurllinklabel}

\bibitem[{{Wegg} \& {Bode}(2010)}]{clus-smbh-TidalDisruptionRates-WeggBode2011}
{Wegg}, C. \& {Bode}, J.~N. 2010, arXiv:1011.5874
 \href{http://xxx.lanl.gov/abs/arXiv:1011.5874}{\urllinklabel}


\bibitem[{{Yardley} {et~al.}(2010){Yardley}, {Hobbs}, {Jenet}, {Verbiest},
  {Wen}, {Manchester}, {Coles}, {van Straten}, {Bailes}, {Bhat},
  {Burke-Spolaor}, {Champion}, {Hotan}, \& {Sarkissian}}]{2010MNRAS.407..669Y}
{Yardley}, D.~R.~B., {Hobbs}, G.~B., {Jenet}, F.~A., {Verbiest}, J.~P.~W.,
  {Wen}, Z.~L., {Manchester}, R.~N., {Coles}, W.~A., {van Straten}, W.,
  {Bailes}, M., {Bhat}, N.~D.~R., {Burke-Spolaor}, S., {Champion}, D.~J.,
  {Hotan}, A.~W., \& {Sarkissian}, J.~M. 2010, \mnras, 407, 669

 \href{http://adsabs.harvard.edu/abs/2010MNRAS.407..669Y}{\adsurllinklabel}

\end{thebibliography}

\end{document}